\documentclass[useAMS,usenatbib]{mn2e}

\usepackage{epsfig,amsmath,fleqn,amssymb}

\title[On an excitation mechanism for trapped inertial waves]{On an excitation mechanism for trapped inertial waves in discs around black holes}
\author[B\'arbara T. Ferreira and Gordon I. Ogilvie]{B\'arbara T. Ferreira$^{1}$\thanks{E-mail:
B.T.Ferreira@damtp.cam.ac.uk} and Gordon I. Ogilvie$^{1}$\\
$^{1}$Department of Applied Mathematics and Theoretical Physics, University of Cambridge, Wilberforce Road, Cambridge CB3 0WA}
\begin{document}

\date{Accepted 2008 March 7. Received 2008 March 7; in original form 2008 February 28}

\pagerange{\pageref{firstpage}--\pageref{lastpage}} \pubyear{2008}

\maketitle

\label{firstpage}

\begin{abstract}
According to one model, high-frequency quasi-periodic oscillations
(QPOs) can be identified with inertial waves, trapped in the inner
regions of accretion discs around black holes due to relativistic
effects. In order to be detected, their amplitudes need to reach large
enough values via some excitation mechanism. We work out in detail a
non-linear coupling mechanism suggested by Kato, in which a global
warping or eccentricity of the disc has a fundamental role. These
large-scale deformations combine with trapped modes to generate
`intermediate' waves of negative energy that are damped as they
approach either their corotation resonance or the inner edge of the
disc, resulting in amplification of the trapped waves. We determine
the growth rates of the inertial modes, as well as their dependence on
the spin of the black hole and the properties of the disc. Our results
indicate that this coupling mechanism can provide an efficient
excitation of trapped inertial waves, provided the global deformations
reach the inner part of the disc with non-negligible amplitude.
\end{abstract}


\section{Introduction}

In the past two decades, and mainly thanks to NASA's \emph{Rossi X-ray
Timing Explorer} (RXTE), about $20$ X-ray binaries, believed to
contain a black hole, have been analysed in detail \citep{bhbbook}. An
important characteristic of these objects is rapid X-ray variability
\citep{klisbook}, in which quasi-periodic oscillations (QPOs) are
included. Of particular relevance are high-frequency quasi-periodic
oscillations (HFQPOs), features observed in the power spectra of the
light curves of black hole candidates, which are a potentially
important tool in the study of strong gravitational fields. The
frequencies observed, and their stability against luminosity
variations, suggest a connection with the inner accretion flow
\citep{remimcclin2006}. Some authors
\citep{nowaketal1997,reldisko,nowaklehrchapter,rkato2001} argued that
these oscillations can be explained in terms of modes arising in the
accretion disc that surrounds the black hole.
Turbulent viscosity, characteristic of accretion flows, generally prevents waves from propagating across the disc to form coherent global modes.
A possible way for such modes to exist
is for them to be trapped in a small region in the
inner part of the disc. Radial trapping of waves was first predicted
by \cite{katofukue1980}, who considered oscillations in a disc around
a Schwarzschild black hole, and was studied further by
\cite{okazakietal1987}. Although trapped waves may reveal relatively
little about the properties of the accretion disc itself, they are a
promising tool in the study of the central object. According to
\cite{remimcclin2006}, a good model of HFQPOs would provide the most
reliable avenue for measuring black hole spins. Once their masses are
also determined, an important step in testing the Kerr metric will be
taken.

For trapped oscillations to explain HFQPOs, an excitation mechanism
for these modes is required, as their amplitudes need to reach values
high enough to allow detection. A possibility can reside in the
interaction between waves in the disc and a global deformation
(warping or eccentricity).  \cite{goodman1993} initially studied the
local excitation of waves in a tidally distorted disc via the
parametric instability.  The idea of using a warp as an excitation
mechanism for disc oscillations goes back to \cite{papterquem1995},
who mentioned the possibility of parametric generation of inertial
waves.  Detailed calculations are reported by
\cite{gammieetal2000}. More recently, and using a different approach,
this problem was studied analytically for thin, relativistic discs
with non-rotating central objects by \cite{katowarp2004} (see also
\cite{kato2007}, where similar calculations are made for eccentric
discs). He made simple estimates for the growth rates of trapped
inertial modes, which are of considerable interest.  On the
other hand, there are many uncertainties in his calculations and he
did not discuss the origin or nature of the global deformations.

In this paper we develop and generalize Kato's ideas on this
excitation mechanism and make detailed numerical calculations of the
modes and growth rates for rotating black holes.  We include a
dynamical treatment of the warp or eccentricity but defer to a second
paper a broader discussion of the origin and global
propagation of these deformations.

In Section 2 we review the trapping of inertial oscillations in a
simple, pseudo-relativistic disc model. In Section 3 we describe the
excitation mechanism for trapped inertial modes, which relies on a
coupling between waves in the disc and global deformations. In Section
4 we discuss the dependence of the inertial modes' growth rates on
disc parameters and black hole spin. Conclusions are presented in
Section 5.

\section[trapping]{Trapped inertial oscillations}
\label{trapping}

\subsection{Basic equations}

The trapping of oscillations can be easily understood by analysing the
fluid equations in a simple isothermal disc model
\citep{lubowpringle1993,rkato2001}. Although this trapping
happens only in discs around compact objects, a fully relativistic
model is not necessary. The most important effects can be
included by supplementing a Newtonian treatment with the correct relativistic
expressions for the characteristic frequencies in the disc
\citep{rkato2001}. For simplicity and clarity, we adopt this
pseudo-relativistic approach and consider a strictly isothermal disc
with a ratio of specific heats $\gamma=1$. Ignoring viscosity and
magnetic fields, the fundamental hydrodynamic equations can be written
as
\begin{equation}
\frac{\partial\bmath{u}}{\partial t}+\bmath{u}\cdot\nabla\bmath{u}=-\nabla{h}-\nabla\Phi,
\label{motion1}
\end{equation}
\begin{equation}
\frac{\partial h}{\partial t} +\bmath{u}\cdot\nabla h =-c_\mathrm{s}^2\nabla\cdot\bmath{u},
\label{energy1}
\end{equation}
where $h=c_\mathrm{s}^2\log\rho$ is the enthalpy and $c_\mathrm{s}$ is the constant
sound speed in the disc.  We neglect self-gravitation and consider a
fixed axisymmetric gravitational potential $\Phi(r,z)$, where
$(r,\phi,z)$ are cylindrical polar coordinates.

\subsection{Linearized equations}

As in the case of stars \citep{cd2002}, to study oscillations one
needs to analyse what happens to this system of equations when the
velocity and enthalpy are perturbed: $q=q_0+q'$. The equilibrium state
of the disc is independent of time and azimuth:
$\bmath{u}_0=\bmath{\Omega}\times\bmath{r}=r
\Omega(r)\bmath{e}_\phi$, where $\Omega$ is the angular velocity,
which is independent of $z$ in a strictly isothermal disc, while $h_0(r,z)$ satisfies $\bmath{\nabla}h_0=r\Omega^2\bmath{e}_r-\bmath{\nabla}\Phi$.  Thus the
perturbations acting on it can be written as
\begin{equation}
q'(r,\phi,z,t)=\textrm{Re}\left[\widetilde{q'}(r,z)\exp(\textrm{i}m\phi-\textrm{i}\omega t)\right],
\end{equation}
where $m$ is the azimuthal mode number and $\omega$ is the oscillation
frequency. Dropping, for simplification, the tildes and zeros,
the linearized equations for the perturbed quantities can be written
in the form
\begin{equation}
-\mathrm{i}\hat{\omega}u'_r - 2\Omega u'_\phi=-\frac{\partial h'}{\partial r},
\label{eq1}
\end{equation}
\begin{equation}
-\mathrm{i}\hat{\omega}u'_\phi +\frac{\kappa^2}{2\Omega}u'_r =-\frac{\textrm{i}mh'}{r},
\end{equation}
\begin{equation}
-\mathrm{i}\hat{\omega}u'_z =-\frac{\partial h'}{\partial z},
\end{equation}
\begin{equation}
-\mathrm{i}\hat{\omega}h'-\Omega_z^2zu'_z=-c_\mathrm{s}^2\left[\frac{1}{r}\frac{\partial (ru'_r)}{\partial r}+\frac{\textrm{i}mu'_\phi}{r}+\frac{\partial u'_z}{\partial z}\right],
\label{eq2}
\end{equation}
where $\hat{\omega}=\omega-m\Omega$ is the Doppler-shifted wave frequency, which is zero at the corotation radius, and $\kappa$ and $\Omega_z$ are the epicyclic and vertical frequencies, respectively \citep{galacticdynamics}.   We apply the thin disc approximation ($\partial h/\partial z=-\Omega_z^2z$), and neglect the term $u'_r\partial h/\partial r$ in the last equation. The latter approximation is valid if the radial wavelength for perturbations is smaller than the radial scale on which the enthalpy varies in the basic state. Variables can then be further separated in $r$ and $z$, using \citep{okazakietal1987}

\begin{equation}
(u'_{r},u'_{\phi},h')=\left(u_{r}(r),u_{\phi}(r),h(r)\right)\textrm{He}_n\left(\frac{z}{H}\right),
\end{equation}
\begin{equation}
u'_{z}=u_z(r)\textrm{He}_{n-1}\left(\frac{z}{H}\right),
\end{equation}
where $\textrm{He}_n$ is the modified Hermite polynomial of order $n$ \citep{abramowitzstegun}, with $n=0,1,2,3,\dots$ being the vertical mode number and $H=\sqrt{c_\mathrm{s}/\Omega_z}$ the vertical scaleheight of the disc. (Since $\textrm{He}_{-1}$ is not defined, for $n=0$ we have $u'_z=0$.) It should be noted that this separation of variables is not exact since $H$ depends on $r$ (as described by \cite{rkato2001}, this separation is valid to lowest WKB order; \cite{nowakwagoner1992} use a slowly varying function of $r$ to separate variables). The variation of $H$ with $r$ couples different vertical modes \citep{tanakaetal2002} but this effect is weak when the radial wavelength is short, and we neglect it.  The final set of ordinary differential equations in $r$ for the perturbed quantities reads

\begin{equation}
-\mathrm{i}\hat{\omega}u_r - 2\Omega u_\phi=-\frac{\mathrm{d}h}{\mathrm{d}r},
\label{free1}
\end{equation}
\begin{equation}
-\mathrm{i}\hat{\omega}u_\phi +\frac{\kappa^2}{2\Omega}u_r =-\frac{\mathrm{i}mh}{r},
\end{equation}
\begin{equation}
-\mathrm{i}\hat{\omega}u_z =-n\frac{h}{H},
\end{equation}
\begin{equation}
-\mathrm{i}\hat{\omega}h-\Omega_z^2Hu_z=-c_\mathrm{s}^2\left[\frac{1}{r}\frac{\mathrm{d} (ru_r)}{\mathrm{d} r}+\frac{\mathrm{i}mu_\phi}{r}\right].
\label{free2}
\end{equation}

\subsection{Wave modes}

The dispersion relation for wave modes in the disc can be determined by further assuming that the radial wavelength of the perturbed quantities is much smaller than both the azimuthal wavelength and the characteristic scale for radial variations of the equilibrium quantities. It can then be verified that perturbations with local radial wavenumber $k$ obey \citep{okazakietal1987}
\begin{equation}
k^2=\frac{(\hat{\omega}^2-\kappa^2)(\hat{\omega}^2-n\Omega_z^2)}{\hat{\omega}^2c_\mathrm{s}^2}.
\label{disprelation}
\end{equation}
As argued before, the most important relativistic effects on wave propagation can be included by using relativistic expressions for the characteristic frequencies. For a particle orbit, these read \citep{kato1990},
\begin{equation}
\Omega=(r^{3/2}+a)^{-1},
\label{relomega}
\end{equation}
\begin{equation}
\kappa=\Omega\sqrt{1-\frac{6}{r}+\frac{8a}{r^{3/2}}-\frac{3a^2}{r^2}},
\label{relkappa}
\end{equation}
\begin{equation}
\Omega_z=\Omega\sqrt{1-\frac{4a}{r^{3/2}}+\frac{3a^2}{r^2}},
\label{relomegaz}
\end{equation}
where $a$ is the dimensionless spin parameter of the central object
($-1<a<1$), $r$ is in units of the gravitational radius $r_{\textrm{g}}=GM/c^2$
and the frequencies are in units of $c^3/GM$, where $c$ is the speed
of light, $G$ the gravitational constant and $M$ the mass of the black
hole. We assume that the characteristic frequencies in the disc can be
approximated by these particle orbit expressions. Of great importance
is the variation of the epicyclic frequency $\kappa$ with $r$: as $r$
decreases, $\kappa$ increases, reaches a maximum and then goes to zero
at the radius of the marginally stable orbit, $r_{\rm ms}$, which can
be regarded as the inner edge of the accretion disc. Since we are
interested in studying oscillations in accretion discs around compact
objects, we adopt these relativistic expressions.  They correctly
describe the frequency and stability of orbits in the Kerr metric, as
well as the apsidal and nodal precession rates, but some information
about the metric coefficients is lost in this pseudo-relativistic
approach.

From the dispersion relation we can see that, if $n\neq0$, two types
of wave-like solutions, propagating in different regions in the disc,
are possible: a high-frequency one with
$\hat{\omega}^2>\textrm{max}(\kappa^2,n\Omega_z^2)=n\Omega_z^2$
(p~mode) and a low-frequency one with
$\hat{\omega}^2<\textrm{min}(\kappa^2,n\Omega_z^2)=\kappa^2$
(r~mode). If $n=0$ the dispersion relation for the non-trivial mode
becomes $\hat{\omega}^2=k^2c_\mathrm{s}^2+\kappa^2$ and waves can
propagate where $\hat{\omega}^2>\kappa^2$; this is the
inertial-acoustic mode. In non-isothermal discs the closest equivalent
of this mode behaves like a surface gravity wave or stellar f~mode
\citep{ogilvie1998,lubowogilvie1998} but here we refer to it
as the $n=0$ mode or 2D mode, since it involves a purely horizontal
motion independent of $z$.

The inertial or r~modes \citep[so called by][]{korycanskypringle1995} are nearly
incompressible since they are restored by inertial forces, avoiding
acoustic effects.  (They are often called g~modes in the literature,
but they are not related to internal gravity waves or stellar
g~modes.)  Acoustic or p~modes have pressure as their main restoring
force and are essentially compressible.  In an isothermal disc, the 2D
mode is a purely horizontal compressible mode.

The propagation regions of the p and r~modes allow us to define important radii in the disc, the resonant radii \citep[e.g.][]{lubowogilvie1998}. Non-axisymmetric waves can have three types of resonances: corotation where $\hat{\omega}=0$, Lindblad resonances where $\hat{\omega}^2-\kappa^2=0$, and vertical resonances where $\hat{\omega}^2-n\Omega_z^2=0$ (for $n\neq 0$).  Lindblad resonances are turning points for the r and 2D modes, while vertical resonances are turning points for p~modes.

The three different types of modes propagate in different regions in the disc. If the epicyclic frequency has a maximum at some particular radius, r and 2D modes can be trapped in the inner part of the disc, while p~modes always propagate to the outer boundary beyond the outer vertical resonance. Originally \cite{katofukue1980} considered the trapping of 2D modes in the very inner region of the disc. These modes can be trapped between the radius of the marginally stable orbit and the inner Lindblad resonance. However, the conditions at $r_\mathrm{ms}$ are not well understood and it is not clear if this trapping region can work as a resonant cavity. Therefore, here we focus on the trapping of r~modes, which happens below the maximum of the epicyclic frequency between two Lindblad resonances (and with no corotation resonance in between), a resonant cavity naturally created by the non-monotonic variation of $\kappa$ with radius \citep{okazakietal1987}.

Of particular importance is the axisymmetric trapped wave with
frequency $\omega\approx\textrm{max}(\kappa)$, and with the simplest
possible radial structure. This mode is trapped in a small region
close to the maximum of the epicyclic frequency, and is important, not
only because it is naturally confined and therefore
more likely to occur in the presence of turbulent viscosity,
but also because its frequency can be identified with $\textrm{max}(\kappa)$, which depends only on the properties of the black hole: its mass $M$ and its angular momentum $a$. Furthermore, this mode is likely to be most easily observed, as it may produce a net luminosity variation of the disc without cancellations. Therefore, measuring the frequency of this mode and determining the mass of the central object by, e.g., studying the orbit of its binary companion, one can, in principle, find the spin of the black hole.

\subsection{Numerical calculation of trapped r~modes}

To find the radial structure of trapped modes we need to solve the
system of equations (\ref{free1})--(\ref{free2}), subject to
appropriate boundary conditions. Numerical calculations of waves
trapped near the maximum of the epicyclic frequency were first
performed by \cite{okazakietal1987}. Here we focus on the simplest
possible trapped inertial modes, with $m=0$ and $n=1$. An approximate
analysis of equations (\ref{free1})--(\ref{free2}) close to the
maximum of the epicyclic frequency shows that, between the two
Lindblad resonances, these solutions are described by parabolic
cylinder functions \citep{abramowitzstegun} involving Hermite
polynomials of order $l=0,1,2,\dots$, centred at the maximum of
$\kappa$, like the solutions of the quantum harmonic oscillator
\citep{perezetal1997}\footnote{It should be noted that
\cite{perezetal1997} use $n$ as the radial mode number. Here the three
quantum numbers are $(l,m,n)$ corresponding to the three coordinates
$(r,\phi,z)$.}. The lowest order mode ($l=0$) has a Gaussian structure
in $r$. In this section we solve the same problem using numerical
methods and fewer approximations, as a prelude to an analysis of the
non-linear mode couplings that cause these modes to grow.

Since we expect to find trapped modes only for some discrete values of the oscillation frequency, we solve the set of equations as a generalized eigenvalue problem, of the form
\begin{equation}
\mathbf{A}\bmath{U}=-\textrm{i}\omega\mathbf{B}\bmath{U},
\end{equation}
where $\bmath{U}$ is the column vector whose components are the r~mode quantities $(u_r,u_\phi,u_z,h)$ evaluated at a set of discrete points, $\mathbf{A}$ is the matrix representing the system of equations (\ref{free1})--(\ref{free2}), and $\mathbf{B}$ can be different from the identity matrix depending on the boundary conditions used. To solve this problem numerically, we use a pseudo-spectral method with Chebyshev polynomials. We use a Gauss--Lobatto grid, $x(i)=\textrm{cos}(\pi i /N)$, where $N$ is the number of grid points, and the Chebyshev coordinate $-1 < x < 1$ is related to the radial coordinate by
\begin{equation}
r=-x\frac{r_\mathrm{out}-r_\mathrm{in}}{2}+\frac{r_\mathrm{out}+r_\mathrm{in}}{2},
\end{equation}
so that $r_\mathrm{in} < r < r_\mathrm{out}$, where $r_\mathrm{in}=r_\mathrm{ms}$ and $r_\mathrm{out}$ is
an outer radius chosen to be larger than the outer Lindblad resonance
for the r~mode.  The representation of the first derivatives in the
matrix $\mathbf{A}$ is achieved using the Chebyshev collocation derivative matrix,
defined by \citep{boydbook}
\begin{equation}
D_{ij}=\begin{cases}
(1+2N^2)/6 & i=j=0 \\
-(1+2N^2)/6 & i=j=N \\ 
-x_j/[2(1-x_j^2)] & i=j, \, j\neq0, N \\
(-1)^{i+j}p_i/[p_j(x_i-x_j)] & i\neq j, 
\end{cases} 
\end{equation}
where $p_0=p_N=2$, and $p_j=1$ otherwise. The generalized eigenvalues and eigenfunctions of $\mathbf{A}$ are then calculated numerically, using IDL's eigenvalue solver, LA\_\textrm{E}IGENPROBLEM, which uses a QR decomposition, and is based on LAPACK routines. The boundary conditions used are the following:
\begin{itemize}
\item {\textbf{At $\bmath{r_\mathrm{in}}$}}$\quad$ $u_{r}=0$. According to the dispersion relation, the r~mode is expected to be exponentially decaying for radii smaller than its innermost Lindblad resonance and therefore its velocity is supposed to be approximately zero at the marginally stable orbit, which justifies the choice of this boundary condition.
\vspace{5pt}
\item {\textbf{At $\bmath{r_\mathrm{out}}$}}$\quad$ $\textrm{d}u_{r}/\textrm{d}r=\textrm{i}ku_{r}$, where $k$ is given by the dispersion relation (\ref{disprelation}) for $m=0$, $n=1$ at $r_\mathrm{out}$, and using $\omega\approx\textrm{max}(\kappa)$. Since the r~mode can propagate again as a p~mode outside the vertical resonance, we choose an outer radius beyond this location, so that the mode is oscillatory there, i.e., $k$ is real, and we select the outgoing wave solution by choosing $k>0$ so that the group velocity of the waves at $r_\mathrm{out}$ is positive.  This condition allows the wave to lose energy through the outer boundary and minimizes artificial wave reflection there.
\end{itemize}

We allow for the possibility that $\omega$ is complex, in which case
its imaginary part is the growth rate of the disturbance.  In fact, in the absence of non-linear mode couplings, we obtain slowly decaying solutions with $\mathrm{Im}(\omega)<0$ as a result of the outgoing-wave outer boundary condition.

\begin{figure}
\includegraphics[width=84mm]{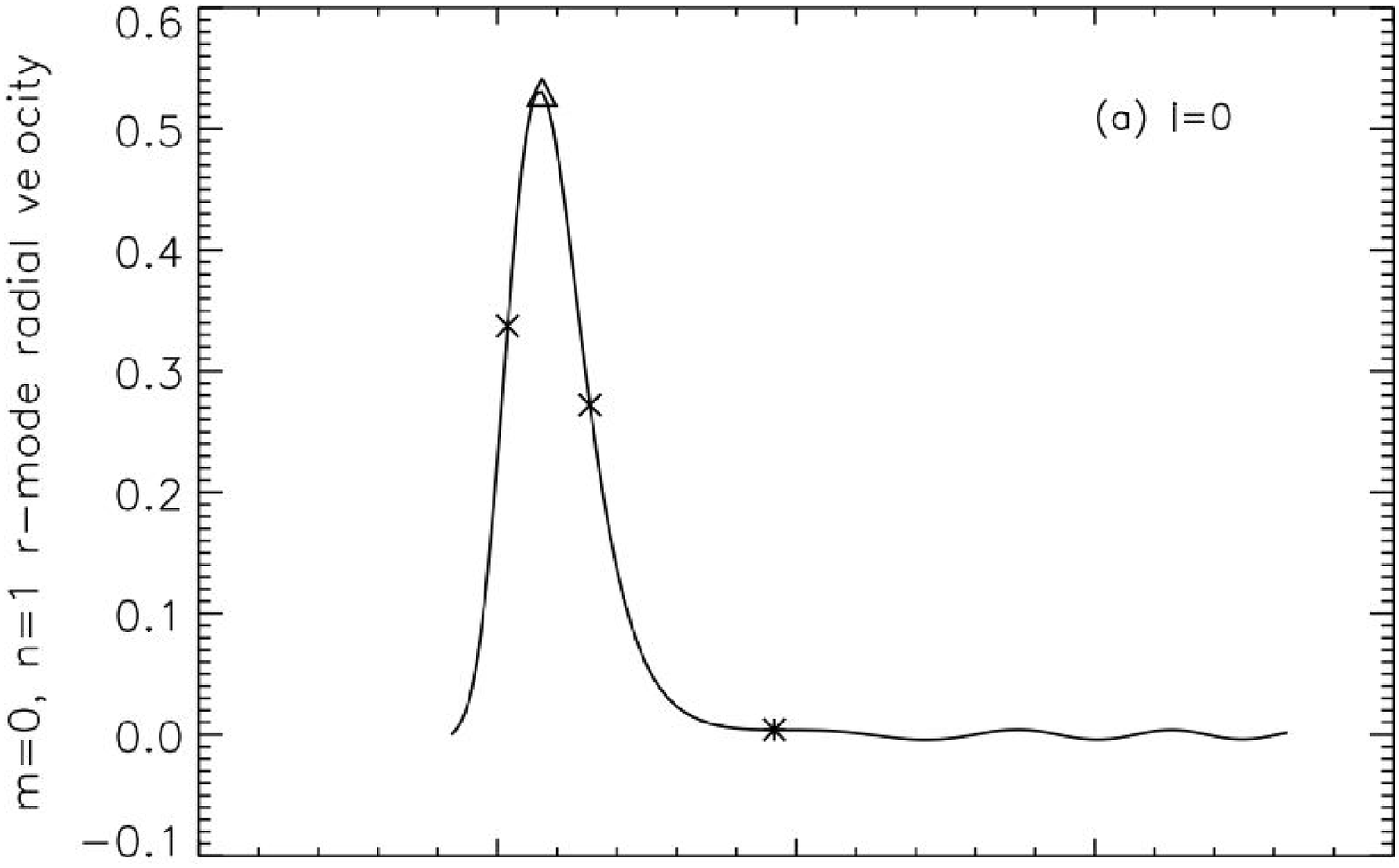}
\includegraphics[width=84.4mm]{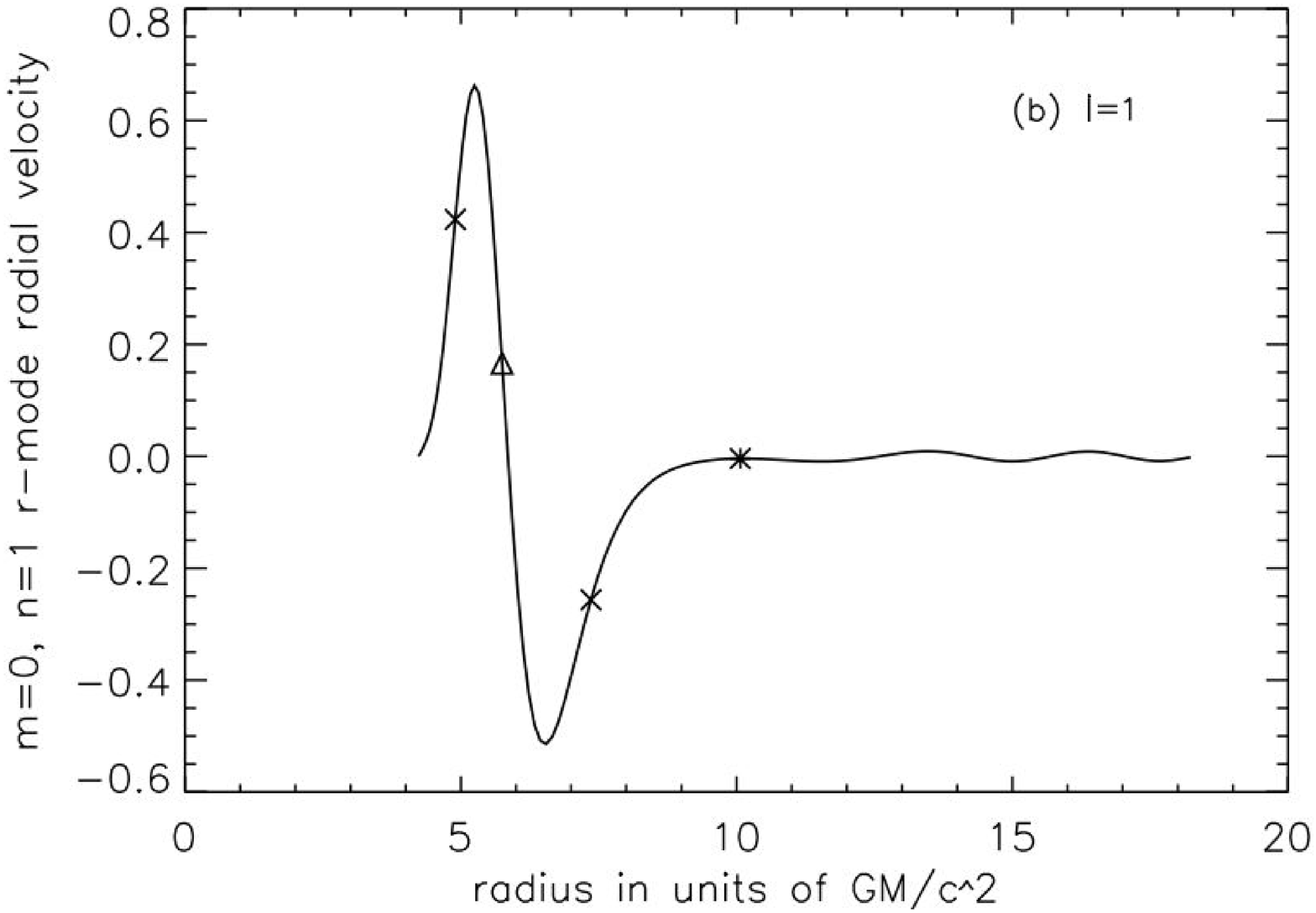}
\caption{Variation of the real part of the radial component of the axisymmetric, $n=1$ r~mode velocity with radius for $c_\mathrm{s}/c=0.01$ and $a=0.5$ for (a) $l=0$ and (b) $l=1$. The triangle indicates the radius where the epicyclic frequency is maximum, and the crosses and asterisks the Lindblad and vertical resonances, respectively.}
\label{freermode}
\end{figure}

In Fig. \ref{freermode} we show the variation of $u_r$ with radius for
two typical trapped solutions, corresponding to two different radial
mode numbers $l=0$ and $l=1$. (Since we are solving an eigenvalue
problem, solutions are multiplied by an arbitrary amplitude.) The
complex frequencies of the modes represented in Fig. \ref{freermode}
are $0.03196-3.6884\times10^ {-7}\textrm{i}$ and
$0.02989-1.1846\times10^ {-7}\textrm{i}$ (in units of $c^3/GM$),
respectively. In these units, and for the value of $a$ used, the
maximum of $\kappa$ is 0.03312. These modes are slightly damped because of the boundary
condition used at the outer radius, which selects the outgoing wave
only. If the sound speed is smaller, the frequency of the modes is
closer to the value of the epicyclic frequency at its maximum, and the
trapping region and damping rate are smaller (see
Tab. \ref{tablefree}).  We obtained identical results by solving
the two-point boundary-value problem using a shooting method.

\begin{table}
\begin{center}
\begin{tabular}{cc}
\hline
$c_\mathrm{s}/c$ & frequency \\
\hline
$0.002$ & $0.03289+0.0\textrm{i}$ \\
$0.005$ & $0.03254-1.276 \times 10^{-10}\textrm{i}$ \\
$0.01$ & $0.03196-3.688 \times 10^{-7}\textrm{i}$ \\
$0.02$ & $0.03079-2.417 \times 10^{-5}\textrm{i}$ \\
\hline
\end{tabular} 
\end{center}
\caption{Dependence of the real and imaginary part of the $l=0$ r~mode frequency (in units of $c^3/GM$) on the sound speed $c_\mathrm{s}$, for $a=0.5$. We use 150 collocation points and a value of 18.2331 for the outer radius, in units of $GM/c^2$ ($r_\mathrm{in}=r_\mathrm{ms}=4.2331$).}
\label{tablefree}
\end{table} 

These inertial modes can be thought of as waves trapped in a virtual potential, $U(r)=-k(r)^2$ \citep{lietal2003}, which for a frequency close to the maximum of $\kappa$ is similar to the harmonic oscillator potential. If $U(r)<0$ waves can propagate, being evanescent in the regions where potential barriers exist. Also, as in quantum mechanics, these trapped inertial waves can escape through the potential barriers and propagate on the other side, as p~modes. The inertial modes are evanescent between the inner radius and the first Lindblad resonance and between the second Lindblad resonance and the vertical one. The `leakage' through the potential barrier is also verified, as our results show small-amplitude oscillations after the vertical resonance (Fig. \ref{freermode}). The larger the sound speed, the larger the width and smaller the height of the barrier, and more `leakage' through the barrier is expected, which is verified numerically (Tab. \ref{tablefree}). This `leakage' was first predicted by \cite{okazakietal1987}.

As described above, the inertial mode characterized by
$(l,m,n)=(0,0,1)$ is likely to be relevant to the interpretation of
observed oscillations. In the next section we describe an excitation
mechanism for this mode, based on non-linear wave coupling.  If the
modes are not strongly coupled, we expect the mechanism to affect
mainly the growth rate $\mathrm{Im}(\omega)$, so that the structure of
the wave still resembles a Gaussian centred at the maximum of
$\kappa$, as shown in Fig. \ref{freermode} (a).

\section[growth]{Growth of oscillations in deformed discs}

Low-frequency modes with azimuthal mode number $m=1$ have been widely
studied in the context of (quasi-)Keplerian accretion disc theory
\citep{kato1983}, since they are global, typically vary on
length-scales comparable to the radial extent of the disc, and are
long-lived. \cite{kato1989} showed that these modes are also global in
a relativistic disc. A global $m=1$ mode with one node in the vertical
direction ($n=1$) is typically identified with a warp in the disc
\citep{papaloizoulin1995}, while $n=0$ modes correspond to eccentric
discs. This is easily seen if we focus on the action of each of these
modes on a ring. The vertical displacement of a $m=1,n=1$ mode is
independent of $z$, and proportional to $\cos(\phi-\textrm{constant})$
at fixed $r$ and $t$, which corresponds to a tilting, as the displacement
\emph{with respect to} the disc plane is different at each azimuthal
angle. In the case of the $n=0$ mode, the radial displacement is the
one that is independent of $z$, and proportional to
$\cos(\phi-\textrm{constant})$ at fixed $r$ and $t$. This means that the
displacement \emph{within} the disc plane varies with $\phi$, creating an
elliptical orbit. As we discuss below, these global deformation modes
can be produced by the presence of a binary companion or by
instabilities.

\subsection{Warped discs}

It is believed that warps exist in many astrophysical discs.
Precessing warped discs have been used successfully to explain
long-term light-curve variations in Her X-1 and some other X-ray
binaries \citep{katz1973,gerendboynton1976}. If the central
object is a compact radiation source, warps can be induced by
radiation pressure forces \citep{pringle1996,wijerspringle1999}.  However, this mechanism is less likely to operate in
systems containing a black-hole primary because the disc needs to be
very large \citep{ogilviedubus2001}.

If the central object is a rotating black hole, its axis of rotation
might not be perpendicular to the plane of the binary orbit in which
the accretion disc forms, in which case the disc is said to be
misaligned or tilted. There is both theoretical and observational
evidence for this tilting (see \cite{globalsimulations} and references
therein). The misalignment of the orbital angular momentum of the disc
and the spin angular momentum of the black hole results in important
changes in the structure of the inner disc as it will be subject to
Lense--Thirring precession \citep{bardeenpetterson1975}. This
differential precession tends to twist the disc, which may adopt a stationary warped shape. Depending on
the `amount of viscosity', the induced warps can propagate
either diffusively, roughly speaking if the \cite{ss73} viscosity parameter
$\alpha$ is greater than $H/r$, or in a wave-like manner if $\alpha$
is smaller than $H/r$.  \cite{ivanovillarionov1997} showed that the warp has an oscillatory radial structure in a low-viscosity disc, and this was investigated further by \cite{lubowetal2002}.

\subsubsection{Variation of disc tilt with radius}

\begin{figure}
\includegraphics[width=84mm]{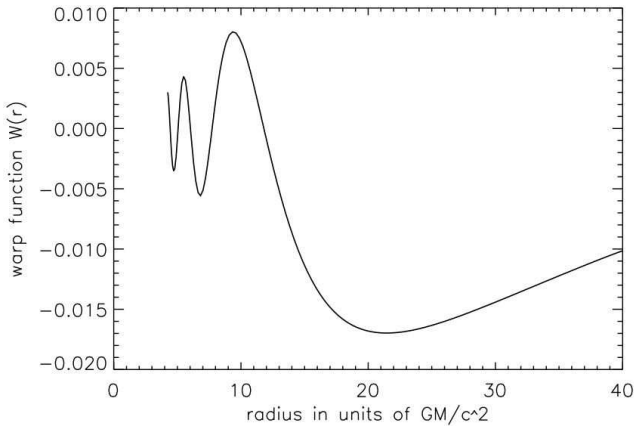}
\caption{Warp function $W(r)$ for $a=0.5$. The sound speed is $0.01c$ and $W(r_\mathrm{in})=W_0=0.003$.}
\label{warpg}
\end{figure}

In a vertically isothermal (pseudo-)relativistic disc, with
$\gamma=1$, a zero-frequency mode with $n=1$, $m=1$ can propagate at
all radii if $a>0$ (i.e.~if the disc and black hole rotate in the same
sense).  This can be seen from the dispersion relation
(\ref{disprelation}) and the expressions (\ref{relkappa}) and
(\ref{relomegaz}), because in this case $\hat\omega^2=\Omega^2$ is
greater than both $\kappa^2$ and $\Omega_z^2$, and therefore $k^2>0$.
Such a stationary warp can be described by linear perturbations of the
form
\begin{equation}
(u'_{\textrm{W}r},u'_{\textrm{W}\phi},h'_\textrm{W})=\left(u_{\textrm{W}r}(r),u_{\textrm{W}\phi}(r),h_\textrm{W}(r)\right) z,
\label{uprime}
\end{equation}
\begin{equation}
u'_{\textrm{W}z}=u_{\textrm{W}z}(r),
\end{equation}
where the subscript W refers to warp quantities, and the
dependence $\textrm{e}^{\textrm{i}\phi}$ is understood. The simplest
possible warp solution is the rigid tilt, valid for a non-rotating
black hole ($a=0$, $\Omega=\Omega_z$), described by $u_{\textrm{W}z}=W\Omega
r$, $u_{\textrm{W}r}=-W\Omega$, $u_{\textrm{W}\phi}=-\mathrm{i}W\textrm{d}(r\Omega)/\textrm{d}r$ and
$h_\textrm{W}=-\mathrm{i}W\Omega^2r$, where $W$ is the constant tilt inclination
(see \cite{papaloizoulin1995} but note that they use $g$ instead of $W$ to represent the disc tilt). If $W$ varies with $r$, equations $(\ref{eq1})$--$(\ref{eq2})$ with $m=1$, $\omega=0$ admit a solution of the
form
\begin{equation}
u'_{\textrm{W}r}=-\Omega Wz+rz\frac{\textrm{d}W}{\textrm{d}r}\frac{\Omega^3}{\Omega^2-\kappa^2},
\end{equation}
\begin{equation}
u'_{\textrm{W}\phi}=-\mathrm{i}Wz\frac{\textrm{d}}{\textrm{d}r}(r\Omega)+\textrm{i}\frac{\Omega\kappa^2}{2(\Omega^2-\kappa^2)}zr\frac{\textrm{d}W}{\textrm{d}r},
\end{equation}
\begin{equation}
u'_{\textrm{W}z}=\Omega r W,
\end{equation}
\begin{equation}
h'_\textrm{W}=-\mathrm{i}\Omega^2Wrz,
\end{equation}
where $W(r)$ is the solution of
\begin{equation}
\frac{\textrm{d}}{\textrm{d}r}\left(\frac{\Omega^2}{\kappa^2-\Omega^2}\frac{\textrm{d}W}{\textrm{d}r}\right)+\frac{1}{r}\frac{\textrm{d}W}{\textrm{d}r}=\frac{\Omega^2-\Omega_z^2}{c_\mathrm{s}^2}W.
\label{eqg}
\end{equation}
This equation\footnote{The above analysis uses the relation
$\kappa^2=4\Omega^2+2r\Omega\,\mathrm{d}\Omega/\mathrm{d}r$, which is
not exactly true of the relativistic expressions because $r^2\Omega$
is not quite the specific angular momentum in relativity. On the other
hand, since the pseudo-relativistic treatment is not fully
self-consistent, if this Newtonian relation is not used here, the
rigid tilt solution is not obtained for $a=0$, contrary to what is
expected physically. Therefore, we choose to use the Newtonian
relation in our treatment of the warp. This is not expected to
significantly influence the final results.}  is closely related, but
not identical, to equation~(17) of \cite{lubowetal2002}, which was
derived from an analysis of global warps in discs that are not
necessarily isothermal.  We solve this equation numerically, using a
4th order Runge--Kutta method with the boundary condition
$\mathrm{d}W/\mathrm{d}r(r_\mathrm{in})=0$, corresponding to zero
torque at the inner edge.  The amplitude of this linear solution may
be fixed by specifying the value $W_0=W(r_\mathrm{in})$ at the inner
boundary, i.e., at the marginally stable orbit; this corresponds to
the (small) inclination of the inner edge of the disc with respect to
the equator of the black hole.  A typical solution is shown in
Fig. \ref{warpg}. The warp has an oscillatory behaviour, as found by
\cite{ivanovillarionov1997}, with the wavelength increasing with
radius, consistent with the local dispersion relation. This
non-monotonic behaviour of the inclination contrasts with the
Bardeen--Petterson effect \citep{bardeenpetterson1975}, which was
derived using an incorrect equation for the warp.
We would normally expect $W(r)$ to tend to a constant value at large
$r$, corresponding to the inclination of the outer part of the disc
with respect to the equator of the black hole.  Unfortunately this is
not true of the approximate equation (\ref{eqg}), which does not hold
accurately at large $r$ because the wavelength becomes comparable to
the radius.  However, since we are interested in the interaction of the
warp with waves that propagate in the inner disc, this is not expected
to significantly affect the final results. We defer to a second paper a more realistic treatment of the propagation of the warp into the inner part of the disc.

\subsubsection{Coupling mechanism}

\begin{figure}
\begin{center}
\includegraphics[width=54mm]{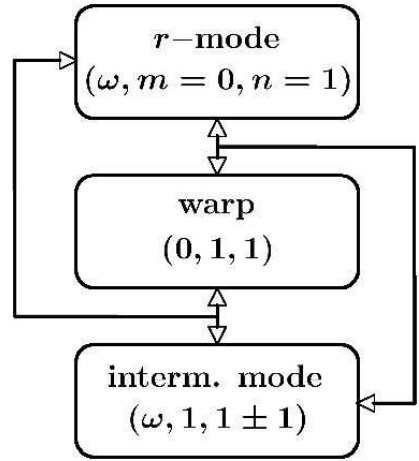}
\end{center}
\caption{Diagram representing the interactions involved in the coupling mechanism involving a warped disc}.
\label{diagram}
\end{figure}

The non-linearities in the basic equations (\ref{motion1}) and
(\ref{energy1}) provide couplings between the different linear modes
of the system.  We are interested in those couplings that lead to
amplification of the trapped modes.  The basic idea of the excitation
mechanism in warped discs \citep{katowarp2004} is that the warp interacts with a wave in
the disc (the trapped r~mode) giving rise to an intermediate
mode. This intermediate mode can then couple with the warp to feed
back on the original oscillations (see Fig. \ref{diagram}), resulting
in growth of the latter.

For the r~mode to be excited, it needs to gain energy in this
coupling. Since the warp has null frequency, its energy is essentially
zero and so the energy exchanges only happen between the r and
the intermediate modes and the disc. It is widely agreed, and certainly true in the short-wavelength limit,
although a general proof is lacking, that a mode that propagates
inside its corotation radius has negative energy, i.e., 
the total energy of the disc is reduced in the presence of the wave, which is possible because the disc is rotating.
On the other hand an axisymmetric wave, such as the r~mode, or one
that propagates outside its corotation radius, has positive energy.
Suppose that, through coupling with the warp, the r~mode
generates an intermediate wave that propagates inside its corotation
radius and therefore has negative energy.  In the process of
generating this wave, the r~mode gains energy and is amplified.  For
sustained growth of the r~mode, the intermediate wave must be damped
so that its negative energy is continually replenished by the r~mode.
(The damping process itself draws positive energy from the rotation of
the disc.)
Therefore a dissipation term should be included in the equations for
the intermediate mode.  We choose to damp this wave locally at a rate $\beta\Omega$, where $\beta$ is a dimensionless parameter.  The origin of this term is not
discussed here but if we interpret it as some type of viscous
dissipation or friction in the disc we expect the intermediate mode,
which propagates in a larger region in the disc, to be more affected
by it than the r~mode, as the latter is trapped in a small
region, and has a simpler radial structure. Also, the intermediate mode approaches its corotation radius (or the marginally stable orbit), where it is expected to be absorbed, and this effect is implicitly included in the intermediate mode equations when the friction term is included. Therefore, we neglect the dissipation term in the equations
for the r~mode. 
The growth rate that we obtain for the trapped mode should be compared with estimates of its damping rate due to turbulent viscosity.

For the coupling to occur the waves need to propagate in the same region in the disc and the parameters $\omega$ and $m$ for the 3 modes need to follow some basic coupling rules,
\begin{equation}
\omega_\textrm{R} \pm \omega_\textrm{W}=\omega_\textrm{I}, \quad m_\textrm{R} \pm m_\textrm{W}=m_\textrm{I},
\label{crules}
\end{equation}
where the subscripts R, W and I refer to r~mode, warp and intermediate mode quantities, respectively.  These rules follow from the quadratic nature of the non-linearitites in the basic equations (\ref{motion1}) and (\ref{energy1}).  Also, we can get information about the vertical mode number $n_\textrm{I}$ of the intermediate mode by remembering that the vertical dependence is given by Hermite polynomials. The warp quantities are proportional to $z$ and the simplest possible r~mode has one node in the vertical direction, therefore its quantities are proportional to $\textrm{He}_1\sim z$. When the warp and this r~mode interact, the coupling terms will be proportional to $z^2/H^2=(z^2/H^2-1)+1\sim\textrm{He}_2+\textrm{He}_0$, i.e., they give rise to two intermediate modes, one with 2 nodes in the vertical direction, $n_\textrm{I}=2$, and a 2D mode with $n_\textrm{I}=0$.
We consider both possibilities.  Note that the frequency of these intermediate waves is that of the r~mode, and they are present because they are forced through the couplings.  They could not exist as free waves satisfying the boundary conditions at this frequency.

According to rules (\ref{crules}), if $\omega_\textrm{R}=\omega$, then $\omega_\textrm{I}=\omega$. For the azimuthal mode numbers, we have $m_\textrm{I}=m_\textrm{R}\pm1$. By analysing the propagation regions for these modes, and considering $\omega\approx \textrm{max}(\kappa+m\Omega)$, it is possible to conclude that the $n=0$ mode with $m_\textrm{I}=m_\textrm{R}-1$ and frequency $\omega$ does not have any Lindblad resonance, i.e., the point where the wave should be excited, in the disc. On the other hand, the mode with $m_\textrm{R}+1$ and frequency $\omega$ has an inner Lindblad resonance close to the region of propagation of the r~mode. For these reasons, we choose the azimuthal mode number of the intermediate mode to be $m_\textrm{R}+1=1$, if the r~mode is axisymmetric. This mode propagates inside its corotation resonance, while the r~mode
has positive energy.  The presence of an inner Lindblad resonance means both that the intermediate mode attains a larger amplitude than it would in the case of nonresonant forcing, and that the flow of energy is such as to amplify the r~mode.

If the vertical mode number of the intermediate mode is chosen to be 2 instead of 0, the r~mode with $(\omega,m,n)=(\omega,0,1)$ interacts with a $(\omega,1,2)$ intermediate mode. The former propagates where $\omega^2<\kappa^2$, while the latter propagates where $(\omega-\Omega)^2<\kappa^2$, where $\omega$ is slightly less than $\max{(\kappa)}$. The propagation regions overlap close to the maximum of the epicyclic frequency since $\Omega\approx2\kappa$ in this region. 
 
The interaction of the r~mode with the $n=2$ intermediate mode in a
warped disc must be treated carefully because in this case the
intermediate mode propagates between its Lindblad resonances and is
absorbed at the corotation resonance. This is a radius in the disc
where the potential $U(r)=-k^2$ tends to infinity and is
therefore difficult to treat numerically because the
wavelength tends to zero. One way of solving this problem is by
including a relatively strong dissipation term in the equations for
the intermediate mode. In this way the wave excited at the inner
Lindblad resonance is damped before reaching corotation. This also
works for the energy exchanges between the modes and the disc, since
the $n=2$ mode has negative energy in the region where it is damped.

\subsubsection{Results}

\begin{figure*}
\includegraphics[width=160mm]{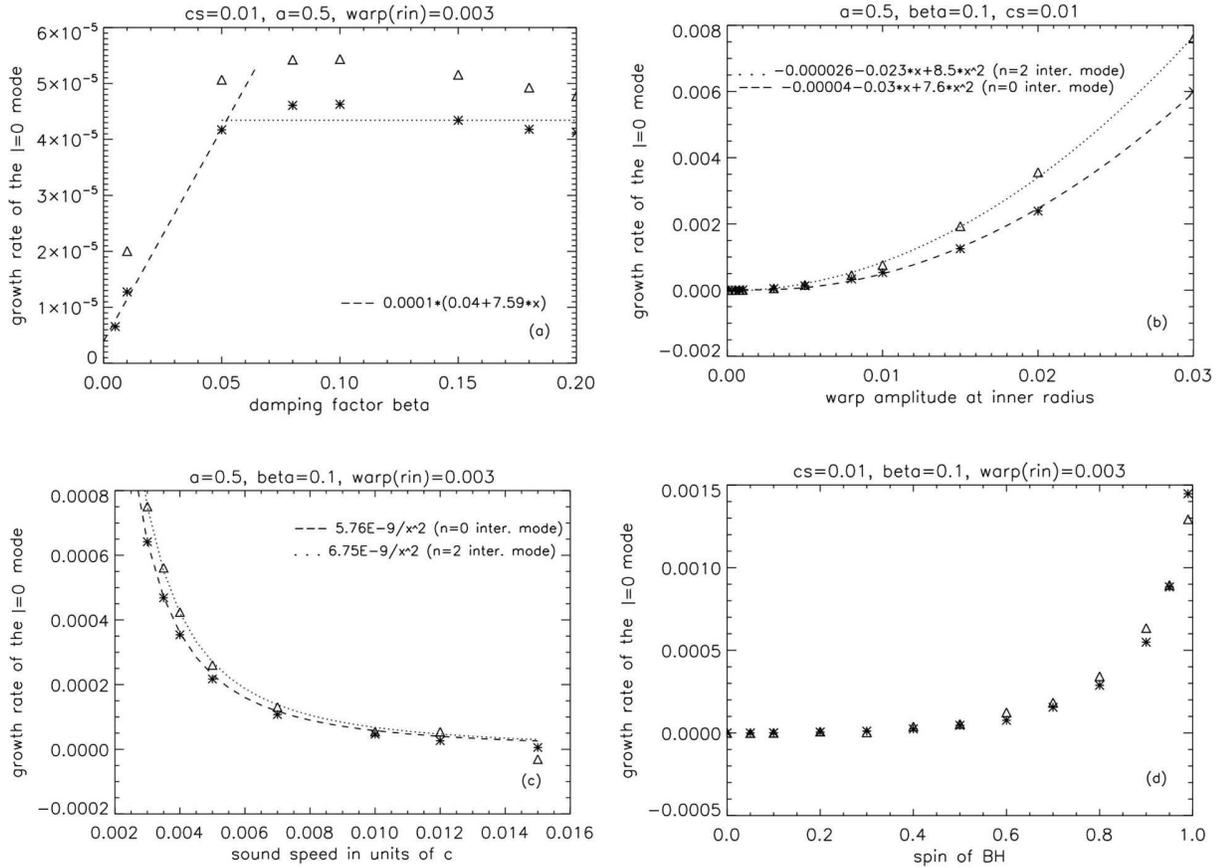}
\caption{Variation of the growth rate of the simplest trapped r~mode, $(l,m,n)=(0,0,1)$, with (a) dissipation factor $\beta$, (b) warp amplitude, $W$, at inner boundary, (c) sound speed in the disc,  and (d) spin of the black hole. The triangles show the results for the interaction with the $n=2$ mode, while the stars are due to the interaction with the $n=0$ mode. For the former, the variation of the growth rate with the dissipation factor is not shown for small values of $\beta$ because of the influence of the corotation resonance in that case.}
\label{relwarp}
\end{figure*}

To find the r~mode growth rate resulting from these interaction we
solve the systems of coupled equations (\ref{fe1})--(\ref{fe2}) and
(\ref{int1})--(\ref{int2}) (see Appendix A) for the interactions of
the r~mode with the $n=0$ and $n=2$ intermediate mode respectively. By
considering the warp to have a fixed amplitude and neglecting the
feedback of the r~mode and intermediate modes on the warp, we still
obtain a linear system of equations, although now the r~mode and
intermediate modes are coupled through the warp.  We treat the $n=0$
and $n=2$ intermediate modes separately, although in practice both
coupling mechanisms act simultaneously and the net growth rate is the
sum of the rates due to the individual mechanisms.

We solve these systems numerically, using the Chebyshev method described in Section \ref{trapping}. For the r~mode the same boundary conditions as before are used. Similar conditions are applied to the intermediate modes, i.e., $u_{\textrm{I}r}=0$ at $r_\mathrm{in}$ and $\textrm{d}u_{\textrm{I}r}/\textrm{d}r=\textrm{i}k_\textrm{I}u_{\textrm{I}r}$ at $r_\mathrm{out}$, where $k_\textrm{I}$ is given by the dispersion relation (\ref{disprelation}) at $r_\mathrm{out}$ for $m=1$ and $n=0$ or $n=2$, depending on the intermediate mode we are considering. As for the r~mode, we choose the sign of $k_\textrm{I}$ so that the outgoing or exponentially decaying wave at $r_\mathrm{out}$ is chosen. The choice of the inner boundary condition for the radial component of the velocity of the $n=2$ mode is justified by the fact that this mode is exponentially decaying there. This choice is harder to justify for the $n=0$ mode since it is oscillatory at $r_\mathrm{in}$. This means that if $u_{\textrm{I}r}=0$ there then the wave is reflected at the inner boundary. Since the conditions at the marginally stable orbit are not clear, we cannot be sure of the physical validity of this condition; we choose it because of its simplicity. Rigorously we would need more boundary conditions to solve this problem, since more than 4 derivatives appear in each system. However, since the coupling terms are expected to be small, $u_{\textrm{R}\phi}$ is roughly proportional to $u_{\textrm{R}r}$ (and similarly for other quantities), thus the boundary conditions imposed for the latter will be indirectly imposed to the former.

\begin{figure}
\includegraphics[width=84mm]{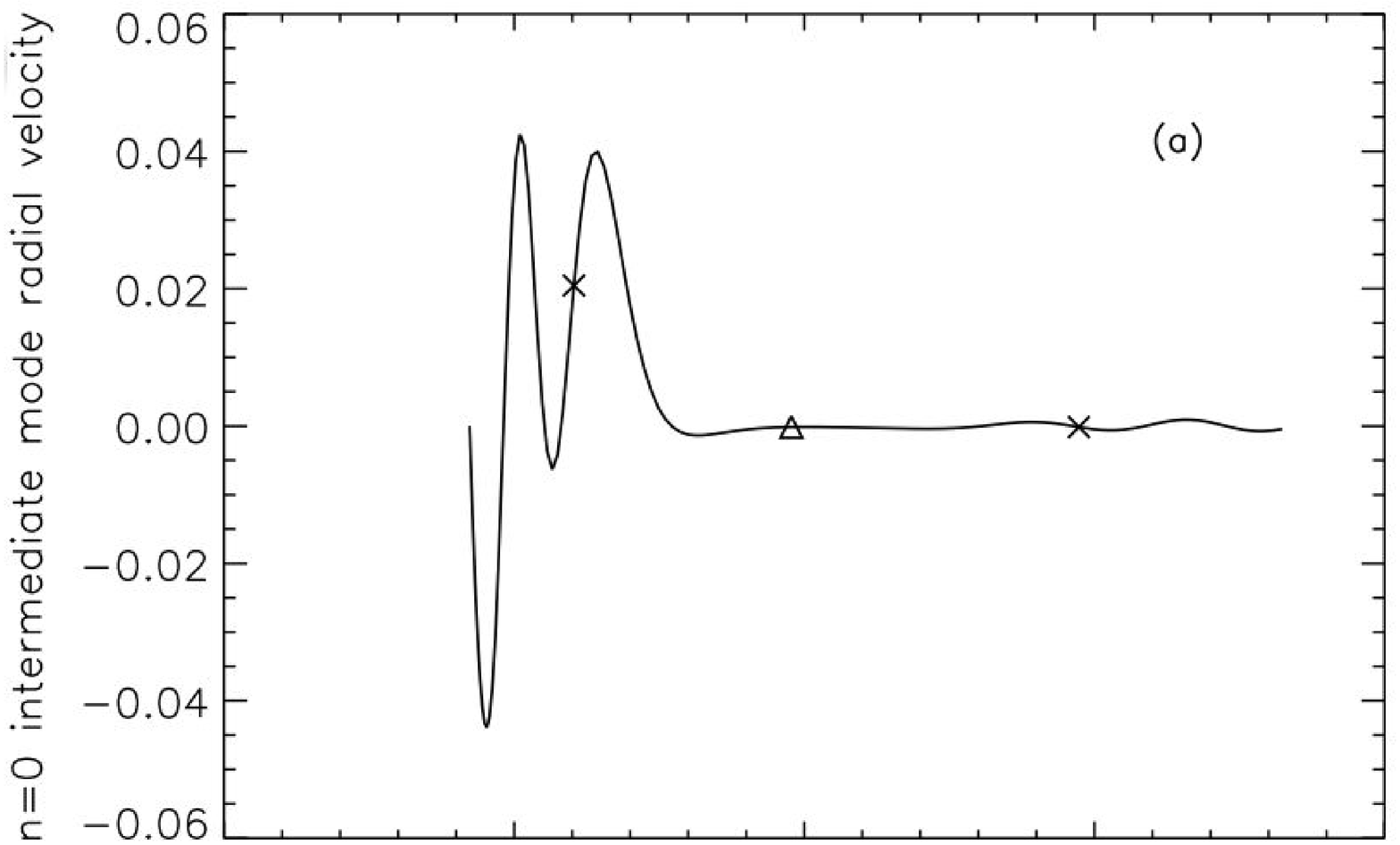}
\includegraphics[width=85mm]{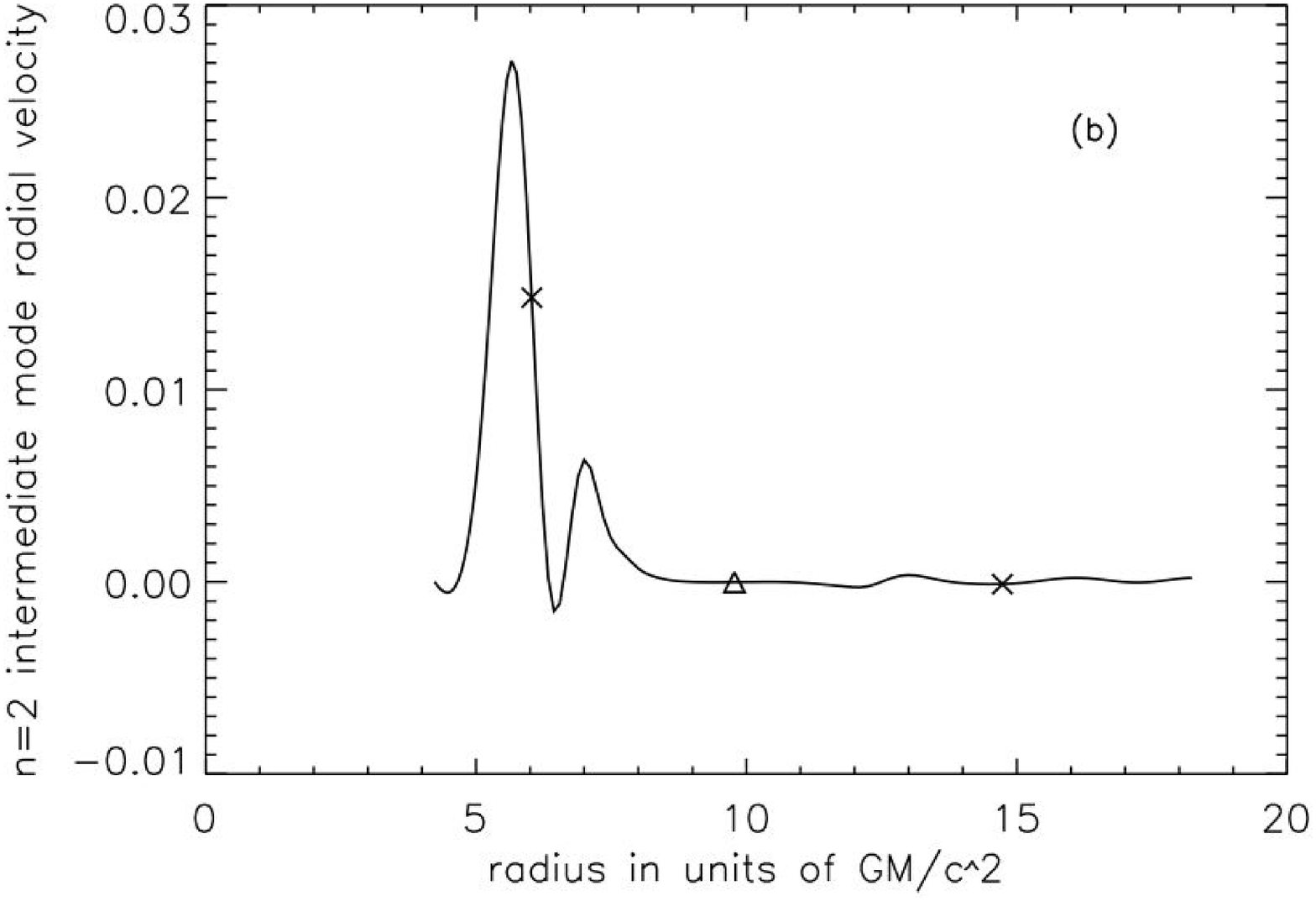}
\caption{Variation of the real part of the radial component of the (a) $m=1$, $n=0$, and (b) $m=1$, $n=2$ intermediate mode velocity with radius for $c_\mathrm{s}/c=0.01$, $a=0.5$, $W(r_\mathrm{in})=0.003$, and $\beta=0.1$. The triangle indicates the radius of the corotation resonance, and the crosses the Lindblad resonances.}
\label{intermediate}
\end{figure} 

The aim is to find the frequencies $\omega$ for which the solutions corresponding to the r~mode are trapped, i.e., for which $u_{\textrm{R}r}$ resembles the parabolic cylinder functions as in Fig. \ref{freermode} (a), which is expected if the coupling terms are small when compared to the other terms in the equations. The imaginary part of $\omega$ then gives the growth rate (or damping rate, if it's negative) of the trapped r~mode. In Fig. \ref{intermediate} we show the $n=0$ and $n=2$ intermediate modes involved in the coupling process, when the dissipation is strong. It should be noted that when the dissipation is weak, the $n=2$ intermediate mode develops a very short wavelength as it approaches the corotation resonance. If $\beta$ is too small, the length-scale on which this wave dissipates is not resolved by our numerical method. The variation of the growth rate with several parameters is shown in Fig. \ref{relwarp}. These results are discussed in Section 4.

\subsection{Eccentric discs}
 
Another possible mechanism for the excitation of trapped waves is
their interaction with an eccentric disc \citep{kato2007}.
Interacting binary stars with mass ratio $q\la0.3$ are believed to
have eccentric accretion discs.  This phenomenon is well documented in
the case of cataclysmic variable stars, where superhumps are observed
during the superoutbursts of the SU UMa class of dwarf novae \citep{pattersonetal2005} and in other
systems of low mass ratio.  In these systems, a resonant interaction
of the orbiting gas with the tidal potential of the companion star
allows a growth of eccentricity \citep{whitehurst1988,lubow1991a,lubow1991b}. Superhumps are also observed in an increasing
number of low-mass X-ray binaries, and systems exhibiting black-hole
HFQPOs are likely to have mass ratios $q\la0.3$ and therefore to have
eccentric discs during at least some phases of their outbursts.

Recently, \cite{kato2007} argued that one-armed global oscillations, symmetric with respect to the $z=0$ plane, can excite trapped oscillations. His conclusions are based on analytical, Lagrangian calculations and are too crude to allow for more than simple estimates for the growth rates. In this section we describe an excitation mechanism similar to the one reported previously, but where an $(m=1,n=0)$ eccentric mode has the role that previously belonged to the $(m=1,n=1)$ warp wave. Using the same numerical method as before, we calculate the trapped r~mode growth rates.

\subsubsection{Variation of eccentricity with radius}
 
\begin{figure}
\includegraphics[width=84mm]{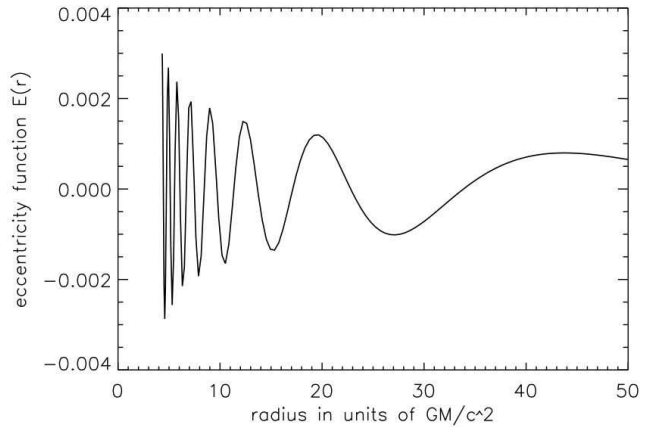}
\caption{Eccentricity function $E(r)$ for $a=0.5$. The sound speed is $0.01c$ and $E(r_\mathrm{in})=E_0=0.003$.}
\label{ecce}
\end{figure}

 As before, consider the set of equations (\ref{free1})--(\ref{free2}). A global eccentric mode corresponds to a zero-frequency wave with $m=1$ and $n=0$.  (If the global eccentric mode precesses freely, the frequency is not exactly zero but is completely negligible compared to the characteristic frequencies in the inner part of the disc.)  In this case, equations (\ref{free1})--(\ref{free2}) are, after the usual separation of variables, reduced to
\begin{equation}
\textrm{i}\Omega u_{\textrm{E}r}-2\Omega u_{\textrm{E}\phi}=-\frac{\textrm{d}h_\textrm{E}}{\textrm{d}r},
\end{equation}
\begin{equation}
\textrm{i}\Omega u_{\textrm{E}\phi}+\frac{\kappa^2}{2\Omega}=-\textrm{i}\frac{h_\textrm{E}}{r},
\end{equation}
\begin{equation}
\textrm{i}\Omega u_{\textrm{E}z}=0,
\end{equation}
\begin{equation}
\textrm{i}\Omega h_\textrm{E}=-c_\mathrm{s}^2\left[\frac{1}{r}\frac{\textrm{d}}{\textrm{d}r}(ru_{\textrm{E}r})+\textrm{i}\frac{u_{\textrm{E}\phi}}{r}\right],
\end{equation}
where the subscript E refers to eccentric mode quantities. This system admits a solution of the form
\begin{equation}
u_{\textrm{E}r}=\textrm{i}E\Omega r,
\end{equation}
\begin{equation}
u_{\textrm{E}\phi}=\frac{c_\mathrm{s}^2 r^2 \Omega}{\Omega^2 r^2 - c_\mathrm{s}^2}\frac{\textrm{d}E}{\textrm{d}r}-\frac{\kappa^2}{2}rE,
\end{equation}
\begin{equation}
u_{\textrm{E}z}=0,
\end{equation}
\begin{equation}
h_\textrm{E}=-\frac{c_\mathrm{s}^2r^2\Omega}{\Omega^2r^2-c_\mathrm{s}^2}\frac{\textrm{d}E}{\textrm{d}r},
\end{equation}
where $E(r)$ is the eccentricity of the disc at radius $r$, and satisfies
\begin{equation}
(\kappa^2-\Omega^2)E=\frac{1}{r^3}\frac{\textrm{d}}{\textrm{d}r}\left(\frac{r^5c_\mathrm{s}^2\Omega^2}{\Omega^2r^2-c_\mathrm{s}^2}\frac{\textrm{d}E}{\textrm{d}r}\right).
\label{eqe}
\end{equation}
Again, this equation is closely related, but not
identical, to equation~(21) of \cite{goodchildogilvie2006},
which was derived from an analysis of global eccentricity in a two-dimensional disc.  Radially propagating solutions are obtained because $\kappa^2<\Omega^2$ in a relativistic disc.
As for the warp tilt $W(r)$, we solve this equation numerically, using a 4th order Runge--Kutta method with boundary conditions $E(r_\mathrm{in})=E_0$, and $\textrm{d}E/\textrm{d}r(r_\mathrm{in})=0$, where $E_0$ is an arbitrary value for the eccentricity at the inner boundary, i.e., at the marginally stable orbit. A typical solution for $E(r)$ is shown in Fig. \ref{ecce}. The eccentricity has an oscillatory
behaviour, with the wavelength decreasing with radius, consistent with the local
dispersion relation. Again, we defer to a second paper a more realistic treatment of the propagation of the eccentricity into the inner part of the disc.

\subsubsection{Coupling mechanism}

The excitation mechanism is similar to the one discussed in the previous section: a global deformation mode, which is now the eccentricity mode, characterized by $(\omega,m,n)=(0,1,0)$, interacts with a trapped r~mode $(\omega,0,1)$ giving rise to an intermediate mode. The latter then couples with the global mode to feedback on to the trapped r~mode.

Coupling rules require the intermediate mode to have the same frequency as the trapped wave and, as before, we have $m_\textrm{I}=1$, in the case where the trapped r~mode is axisymmetric. As for the vertical dependence, since the eccentric mode has $n=0$, the intermediate mode can only have the same vertical mode number as the trapped mode, i.e., $n_\textrm{I}=1$. The propagation region for this mode is the same as for the $(\omega,1,2)$ intermediate mode, present in the interaction of the r~mode with the warp. As in that case, a relatively large damping term must be included in the equations for the intermediate mode so that it dissipates on a resolved scale before reaching the corotation resonance. In this case the energy exchanges are similar to the ones discussed for the interaction in a warped disc.

To find the growth rates that result from this interaction, we solve equations (\ref{inte1})--(\ref{inte2}), using the same numerical method and boundary conditions as before. 

\begin{figure*}
\includegraphics[width=160mm]{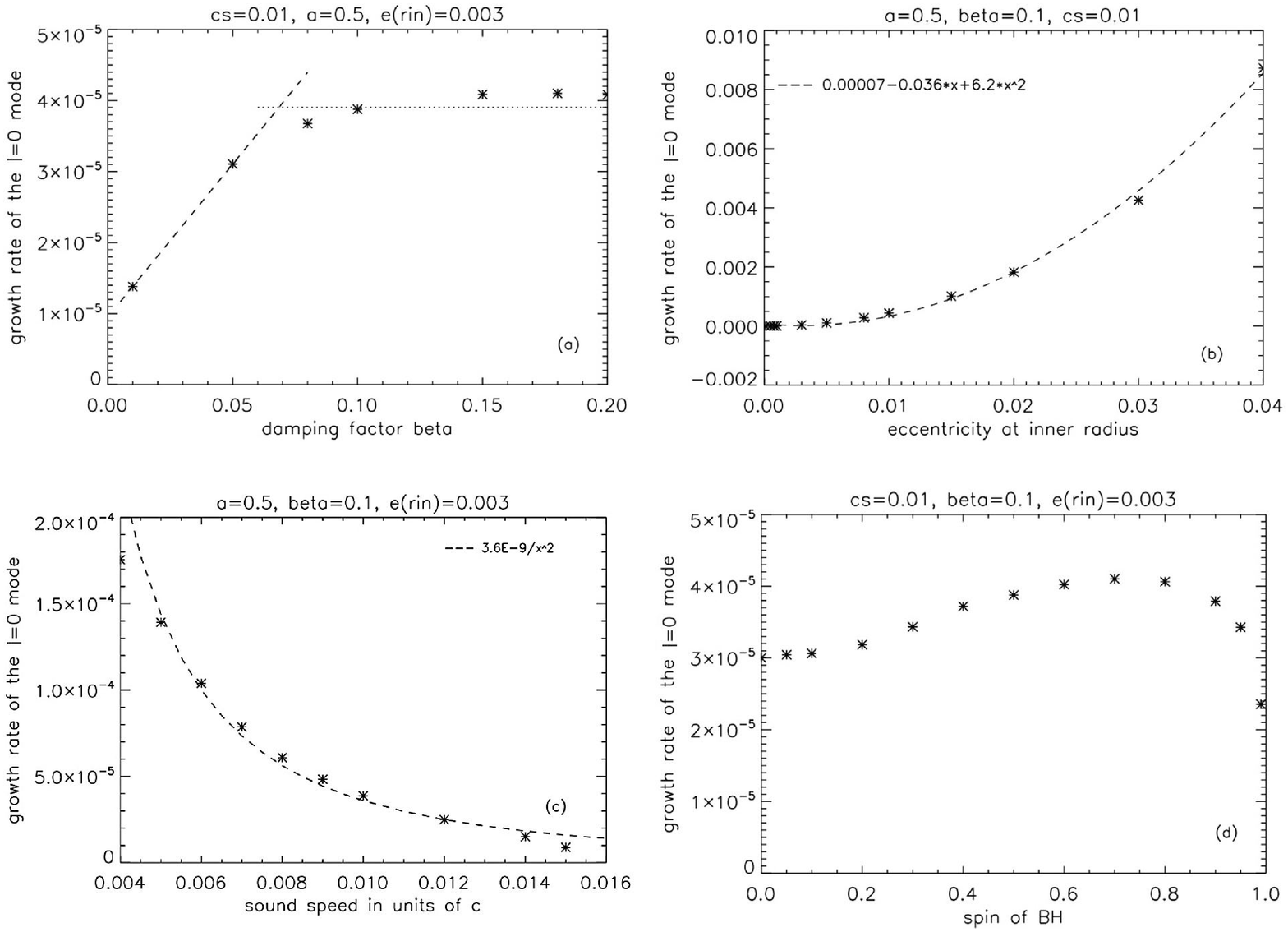}
\caption{Variation of the growth rate of the simplest trapped r~mode,
$(l,m,n)=(0,0,1)$,
 with (a) dissipation factor $\beta$, (b) eccentricity amplitude, $E$, at inner boundary, (c) sound speed of the disc,  and (d) spin of the black hole.}
\label{relecc}
\end{figure*}

The variation of the growth rate with the inner eccentricity, sound speed, spin of black hole and dissipation factor is shown in Fig. \ref{relecc}. The values of the growth rate achieved in the interaction of the trapped wave with the eccentric mode are, in general, similar to the values obtained in the interaction with the warp, if the inner inclination and eccentricity are similar. 

\section{Discussion}

In this section we discuss the results shown in Figs \ref{relwarp} and \ref{relecc}, where the dependence of the $l=0$ r~mode growth rate with several parameters is represented.

\subsection{Growth rates in warped discs}

In the variation of the growth rate with the dissipation factor, two regimes can be considered: weak dissipation ($\beta \lesssim 0.05$), where the variation is approximately linear, and strong dissipation ($\beta \gtrsim 0.05$), where the growth rate remains approximately constant when $\beta$ varies (Fig. \ref{relwarp} (a)). In the former case, the $n=0$ intermediate mode is launched at its inner Lindblad resonance and propagates, being slightly attentuated, until it reaches the inner boundary where it is reflected. Owing to the attenuation, the reflected wave amplitude is smaller than the incident one, and therefore the intermediate mode does not cancel itself, leaving a small amount of energy available for the r~mode to be excited. In the strong dissipation regime, the $n=0$ mode is dissipated before reaching the marginally stable orbit. In this case, all the energy carried by this wave becomes available to excite the trapped mode. As for the $n=2$ intermediate mode, in the strong dissipation regime, the wave is completely dissipated before reaching the corotation resonance. Physically we would expect no dissipation term to be necessary in this case. An arbitrarily small amount of dissipation should lead in principle to the complete absorption of the wave at the corotation resonance. However, this cannot be verified numerically because of the difficulty in resolving the wavelength of the intermediate mode as the singularity at the corotation resonance is approached.

For small warp amplitudes, i.e., before the coupling terms start affecting the structure of the eigenfunctions, the growth rate grows with the square of the warp amplitude at the inner boundary (Fig. \ref{relwarp} (b)). This is expected since the coupling mechanism relies on the `use' of the warp twice: first on the interaction with the r~mode to give rise to the intermediate modes, and then again on the interaction with the latter to feed back on the former (Fig. \ref{diagram}).

The excitation mechanism discussed here is similar to the well known parametric instability, in the case where one of the modes is strongly damped. The parametric instability is a type of resonant coupling between three modes satisfying $\omega_\textrm{p}\approx\omega_{\textrm{d}1}+\omega_{\textrm{d}2}$, where the subscripts p and d refer to parent and daughter modes, respectively. The parametric instability results in the transfer of energy from the former to the latter, when the daughter modes have small amplitude. The equations describing the evolution of the mode amplitudes read \citep[adapted from]{wugoldreich2001},
\begin{equation}
\frac{\textrm{d}A_\textrm{p}}{\textrm{d}t}=+\gamma_\textrm{p}A_\textrm{p}-\textrm{i}\omega_\textrm{p}A_{\textrm{p}}+\textrm{i}\omega_\textrm{p}\sigma A_{\textrm{d}1} A_{\textrm{d}2},
\end{equation}
\begin{equation}
\frac{\textrm{d}A_{\textrm{d}1}}{\textrm{d}t}=-\gamma_{\textrm{d}1}A_{\textrm{d}1}-\textrm{i}\omega_{\textrm{d}1}A_{\textrm{d}1}+\textrm{i}\omega_{\textrm{d}1}\sigma A_{\textrm{p}} A^*_{\textrm{d}2},
\end{equation}
\begin{equation}
\frac{\textrm{d}A_{\textrm{d}2}}{\textrm{d}t}=-\gamma_{\textrm{d}2}A_{\textrm{d}2}-\textrm{i}\omega_{\textrm{d}2}A_{\textrm{d}2}+\textrm{i}\omega_{\textrm{d}2}\sigma A^*_{\textrm{d}1} A_{\textrm{p}},
\end{equation}
where $\gamma_j>0$ is the linear amplitude growth/damping rate of mode $j$ and $\sigma$ is the non-linear coupling constant. Let us consider a simplified case where the amplitude of the parent mode is approximately constant in time (because the daughter modes are of small amplitude), and $\omega_{\textrm{d}1}=-\omega_{\textrm{d}2}=\omega$, $\gamma_{\textrm{p}}=\gamma_{\textrm{d}1}=0$ and $\gamma_{\textrm{d}2}=\gamma$. In this case, the parent mode can be compared to the warp while the daughter modes 1 and 2 can be compared with the r and intermediate modes, respectively. Assuming $A_{\textrm{d}1}\propto\exp{(st)}$, the growth rate is
\begin{equation}
\textrm{Re}(s)=-\frac{\gamma}{2}+\left(\frac{\gamma^2}{4}+|A_{\textrm{p}}|^2\sigma^2\omega^2\right)^{1/2}.
\end{equation}
If
$\gamma\ll|A_\mathrm{p}|\sigma\omega$,
$\textrm{Re}(s)\approx|A_{\textrm{p}}|\sigma\omega-\frac{\gamma}{2}$,
i.e., the growth rate is linearly related to the amplitude of the parent mode. On the other hand, if 
$\gamma\gg|A_\mathrm{p}|\sigma\omega$,
$\textrm{Re}(s)\approx|A_\mathrm{p}|^2\sigma^2\omega^2/\gamma$, i.e., the growth rate is proportional to the square of the amplitude of the parent mode. The latter case is the one similar to the excitation mechanism we are discussing here. It should be noted that this parametric instability analysis gives a dependence of the growth rate in $\gamma$ which is not in agreement with the numerical results (considering $\gamma$ to be equivalent to $\beta$), because the dependence of the spatial structure of the intermediate mode on the dissipation is not considered in this simplistic analysis. Also, the parametric instability analysis suggests that the daughter modes gain energy from the parent mode, which is not what happens in the coupling mechanism we are considering, since here the differential rotation of the disc, and not the warp, is the ultimate source of energy for the r~mode. Although the parametric instability analysis gives a dependence of the growth rate on the disturbance amplitude in agreement with our numerical results, it is simplistic and does not consider all the details of the coupling mechanism. A parallel between the parametric instability and our excitation mechanism is therefore not straightforward.

In the strong dissipation regime, the intermediate mode dissipates completely, and does not influence the variation of the r~mode growth rate with both the sound speed of the disc and the spin of the black hole. In this regime, the growth rate decreases with increasing $c_\mathrm{s}$, as expected. The hotter the disc is, the wider the modes get, which means that if the sound speed is high, the modes are not as well trapped. More importantly, the shape of the warp changes when the sound speed changes since its wavelength ($\lambda_\textrm{W}$) is proportional to $c_\textrm{s}$. The interaction relies on the use of the warp twice, therefore we can argue that the growth rate is proportional to $|dW/dr|^2$ (since $W$ represents the inclination and $dW/dr$ the actual warp). Since $|dW/dr|^2\propto W_0^2/\lambda_\textrm{W}^2\propto W_0^2/c_\textrm{s}^2$, i.e., for fixed sound speed the growth rate is proportional to the square of the inner warp amplitude (Fig. \ref{relwarp} (b))  and for fixed $W_0$, the growth rate varies with $1/c_\textrm{s}^2$ (Fig. \ref{relwarp} (c)). The small changes to the $1/c_\mathrm{s}^2$ law are justified by the fact that, for large sound speed, the decay rate due to the `leakage' at $r_\textrm{out}$ is considerable.

As for the variation of the growth rate with the spin of the black hole, an important conclusion is that, in fact, as argued in the beginning of this section, there is no mode excitation if the black hole is non-rotating. This was not evident in Kato's simple estimates for the growth rates. The fact that the growth rate increases with $a$ is also expected, not only because the waves are better confined for larger values of $a$, but mainly because the average warp amplitude in the trapped region is larger, for larger $a$. Also, the wavelength of the warp decreases as $a$ increases, because the Lense--Thirring frequency ($\Omega_{\textrm{LT}}=\Omega-\Omega_z\approx 2a/r^3$) increases. Therefore, for the same inner $W$, a larger $dW/dr$ is achieved. In light of these very preliminary results, we could argue that HFQPOs would preferentially be detected in black hole candidates with large spin, if the interaction of the trapped r~mode with a warped disc is the mechanism responsible for the excitation of the former.

\subsection{Growth rates in eccentric discs}

For the interaction between the r and intermediate modes in an eccentric disc, the variation of the growth rate with the dissipation factor is similar to the same variation for the interaction in a warped disc (Figs \ref{relecc} (a) and \ref{relwarp} (a)). Also, the change of the growth rate with the inner eccentricity (Fig. \ref{relecc} (b)) is very similar to the variation of the growth rate with the warp tilt at the inner radius, i.e., there is a square dependence on the eccentricity amplitude at $r_\mathrm{in}$. This is expected as the eccentric mode plays, in the wave interaction described in this section, the role of the warp in the interactions described previously. Similarly, the variation with the sound speed (Fig. \ref{relecc} (c)) is also the expected one. 

The main difference between the interaction with the warp and the interaction with the eccentric mode is in the variation of the growth rate with the spin of the black hole (Fig. \ref{relecc} (d)). The warp, being a $n=1$ mode, has a variation with radius that strongly depends on the Lense-Thirring precession frequency, and therefore, that strongly depends on $a$. On the other hand, the variation of the eccentricity amplitude with radius, given by equation (\ref{eqe}), is less dependent on the spin of the black hole. Therefore, the variation of $a$ only causes variations of a factor of, at maximum, $2$ in the growth rate obtained for the interaction within the eccentric disc. A very important difference is the fact that, in this interaction, a reasonable growth rate can be obtained in the case where $a=0$. Therefore, in slowly rotating black holes, HFQPOs might be detected if the disc is eccentric, and if this excitation mechanism is responsible for the increase in the amplitude of oscillations.

\section{Conclusions}

In this paper we have described an excitation mechanism for trapped inertial
modes, based on a non-linear coupling mechanism between these waves and global
deformations (warping or eccentricity) in accretion discs. We have seen that
the interaction of a trapped r~mode with an intermediate mode and a deformation
in the disc results in growth of the trapped mode, if there is some process
capable of making the intermediate mode dissipate in the disc. Dissipation is
required so that this mode can remove rotational kinetic energy from the disc,
which becomes available for the r~mode to grow. Depending on the values of the
sound speed, spin of the black hole and amplitude of the deformation at the
inner radius, reasonable growth rates can be obtained for a warp or
eccentricity of modest amplitude. In a warped disc, where the growth rate
varies significantly with the spin of the compact object, growth rates as large
as $\omega/10$, where $\omega$ is the oscillation frequency, can be obtained.
If $a=0$, no oscillations are excited in these discs. However, it may still
possible to excite trapped modes in discs around non-rotating black holes if
they are eccentric.

The coupling process described here works as an excitation mechanism for trapped inertial waves, under a wide range of conditions, provided global deformations reach the inner disc region with non-negligible amplitude. The propagation of global modes, in a more realistic disc model, is the subject of a forthcoming paper \citep{ferreiraogilvie2008}. 

In this paper we considered the excitation of trapped waves due to a non-linear coupling mechanism with global deformations, in a simple disc model. While this effect is responsible for the growth of these modes, it has to compete with others that contribute to the damping of these waves. For example, since the conditions at the marginally stable orbit are unknown, it is possible for a `leakage' of the trapped mode (similar to the one considered in the potential barrier analogy) through $r_\textrm{ms}$ to exist. This effect is to be considered in the future. Also, and more importantly, viscous dissipation in the disc can cause damping of these modes. A simple estimate gives a damping rate of $\alpha\Omega$. For small enough values of $\alpha$ and large enough warp or eccentricity, net growth can occur.

Another point to be discussed is the applicability of our results to observed
discs. We consider a very simple, isothermal disc model and wave perturbations
for which $\gamma=1$. In a more realistic disc, the vertical structure of the
waves is changed while they propagate radially. The wave energy concentrates
either near the surface of the disc \citep{lubowogilvie1998} or towards the
disc mid-plane \citep{korycanskypringle1995}, which could potentially hinder
the propagation of intermediate modes away from the Lindblad resonance where
they are excited. However, the process of `wave channelling' mentioned by
\cite{lubowogilvie1998} is only relevant at a distance from the resonance of
$\sim r_\textrm{L}/m$ ($r_\textrm{L}$ being the radius of the Lindblad
resonance), where the radial wavelength becomes comparable to the semithickness
of the disc. Since the intermediate modes have $m=1$, this effect is not
important in the region where wave coupling occurs. The same is expected for
cases in which the energy concentrates towards the disc mid-plane. The global
deformation modes do not undergo significant wave channelling because their
wavelengths are always long compared to $H$.  Therefore, we believe that our
results, obtained in a simple disc model, are still qualitatively valid in more
realistic discs.

\section*{Acknowledgments}

We thank John Papaloizou for some helpful suggestions, in particular for pointing out the possibility of a relation between the coupling mechanism described here and the parametric instability. We also thank an anonymous referee for useful comments. The work of BTF was supported by FCT (Portugal) through grant no. SFRH/BD/22251/2005.

\newpage
\onecolumn

\appendix

\section[]{Non-linear coupling equations}

\subsection{Wave coupling in a warped disc}

The following system of equations (cf. equations (\ref{motion1})--(\ref{energy1})) describes the propagation of the r~mode and $n=0$ intermediate mode, coupled by the warp, and needs to be solved for the growth rate to be determined:
\begin{equation}
\left(\frac{\partial}{\partial t}+\Omega\frac{\partial}{\partial \phi}\right)u'_{\textrm{R}r}-2\Omega u'_{\textrm{R}\phi}=-\frac{\partial h'_\textrm{R}}{\partial r}+f_{\textrm{R}r},
\label{wqpo1}
\end{equation}
\begin{equation}
\left(\frac{\partial}{\partial t}+\Omega\frac{\partial}{\partial \phi}\right)u'_{\textrm{R}\phi}+\frac{\kappa^2}{2\Omega}u'_{\textrm{R}r}=-\frac{1}{r}\frac{\partial h'_\textrm{R}}{\partial \phi}+f_{\textrm{R}\phi},
\end{equation}
\begin{equation}
\left(\frac{\partial}{\partial t}+\Omega\frac{\partial}{\partial \phi}\right)u'_{\textrm{R}z}=-\frac{\partial h'_\textrm{R}}{\partial z}+f_{\textrm{R}z},
\end{equation}
\begin{equation}
\left(\frac{\partial}{\partial t}+\Omega\frac{\partial}{\partial \phi}\right)h'_\textrm{R}-\Omega_z^2 z u'_{\textrm{R}z}=-c_\mathrm{s}^2\left[\frac{1}{r}\frac{\partial(ru'_{\textrm{R}r})}{\partial r}+\frac{1}{r}\frac{\partial u'_{\textrm{R}\phi}}{\partial\phi}+\frac{\partial u'_{\textrm{R}z}}{\partial z}\right]+f_{\textrm{R}h},
\label{wqpo2}
\end{equation}
\begin{equation}
\left(\frac{\partial}{\partial t}+\Omega\frac{\partial}{\partial \phi}\right)u'_{\textrm{I}r}-2\Omega u'_{\textrm{I}\phi}=-\frac{\partial h'_\textrm{I}}{\partial r}-\beta\Omega u'_{\textrm{I}r}+f_{\textrm{I}r},
\label{wqpo3}
\end{equation}
\begin{equation}
\left(\frac{\partial}{\partial t}+\Omega\frac{\partial}{\partial \phi}\right)u'_{\textrm{I}\phi}+\frac{\kappa^2}{2\Omega}u'_{\textrm{I}r}=-\frac{1}{r}\frac{\partial h'_\textrm{I}}{\partial\phi}-\beta\Omega u'_{\textrm{I}\phi}+f_{\textrm{I}\phi},
\end{equation}
\begin{equation}
\left(\frac{\partial}{\partial t}+\Omega\frac{\partial}{\partial \phi}\right)h'_\textrm{I}=-c_\mathrm{s}^2\left[\frac{1}{r}\frac{\partial(ru'_{\textrm{R}r})}{\partial r}+\frac{1}{r}\frac{\partial u'_{\textrm{I}\phi}}{\partial \phi}\right]-\beta\Omega h'_{\textrm{I}}+f_{\textrm{I}h},
\label{wqpo4}
\end{equation}
where
\begin{equation}
\bmath{f}_\textrm{R}=\left(f_{\textrm{R}r},f_{\textrm{R}\phi},f_{\textrm{R}z}\right)=-\bmath{u}'_\textrm{I}\cdot\nabla\bmath{u}'_\textrm{W}-\bmath{u}'_\textrm{W}\cdot\nabla\bmath{u}'_\textrm{I},
\label{coup1}
\end{equation}
\begin{equation}
f_{\textrm{R}h}=-\bmath{u}'_\textrm{I}\cdot\nabla h'_\textrm{W}-\bmath{u}'_\textrm{W}\cdot\nabla h'_\textrm{I},
\label{coup2}
\end{equation}
\begin{equation}
\bmath{f}_\textrm{I}=\left(f_{\textrm{I}r},f_{\textrm{I}\phi},f_{\textrm{I}z}\right)=-\bmath{u}'_\textrm{R}\cdot\nabla\bmath{u}'_\textrm{W}-\bmath{u}'_\textrm{W}\cdot\nabla\bmath{u}'_\textrm{R},
\label{coup3}
\end{equation}
\begin{equation}
f_{\textrm{I}h}=-\bmath{u}'_\textrm{R}\cdot\nabla h'_\textrm{W}-\bmath{u}'_\textrm{W}\cdot\nabla h'_\textrm{R}
\label{coup4}
\end{equation}
are the coupling terms, arising from non-linearities in the basic equations.

Since we are interested in studying the axisymmetric r~mode, the azimuthal mode number for the intermediate mode is $1$. Also, since the simplest possible r~mode has one node in the vertical direction and the intermediate mode has $n=0$, we use the following separation of variables:

\begin{equation}
(u'_{\textrm{R}r},u'_{\textrm{R}\phi},h'_\textrm{R})=\textrm{Re}\left[\left(u_{\textrm{R}r}(r),u_{\textrm{R}\phi}(r),h_\textrm{R}(r)\right)\textrm{He}_1\left(\frac{z}{H}\right)\textrm{e}^{-\textrm{i}\omega t}\right], 
\end{equation}
\begin{equation}
u'_{\textrm{R}z}=\textrm{Re}\left[u_{\textrm{R}z}(r)\textrm{He}_0\left(\frac{z}{H}\right)\textrm{e}^{-\textrm{i}\omega t}\right],
\end{equation}
\begin{equation}
(u'_{\textrm{I}r},u'_{\textrm{I}\phi},h'_\textrm{I})=\textrm{Re}\left[\left(u_{\textrm{I}r}(r),u_{\textrm{I}\phi}(r),h_\textrm{I}(r)\right)\textrm{He}_0\left(\frac{z}{H}\right)\textrm{e}^{\textrm{i}\phi-\textrm{i}\omega t}\right],
\end{equation}
\begin{equation}
u'_{\textrm{I}z}=0 \quad \textrm{(2D mode)},
\end{equation}
\begin{equation}
(u'_{\textrm{W}r},u'_{\textrm{W}\phi},h'_\textrm{W})=\textrm{Re}\left[\left(u_{\textrm{W}r}(r),u_{\textrm{W}\phi}(r),h_\textrm{W}(r)\right) z\, \textrm{e}^{\textrm{i}\phi}\right],
\end{equation}
\begin{equation}
u'_{\textrm{W}z}=\textrm{Re}\left[u_{\textrm{W}z}(r)\, \textrm{e}^{\textrm{i}\phi}\right].
\end{equation}

This separation of variables results in having the coupling terms (\ref{coup1}) and (\ref{coup2}) proportional to $\textrm{He}_1$ only, while the coupling terms (\ref{coup3})--(\ref{coup4}) give rise to terms proportional to both $\textrm{He}_0=1$ and $\textrm{He}_2=z^2/H^2-1$. Since we are interested in the terms that influence the mode with $n=0$, we need to project these forcing terms on to $\textrm{He}_0$. Also, the separation of variables results in coupling terms of the form
\begin{equation}
\textrm{Re}(A)\textrm{Re}(B)=\frac{1}{2}\textrm{Re}(AB+AB^*),
\end{equation}
which means that the interaction of the $m=0$ r~mode with the $m=1$ warp results in two new modes, one with $m=1$ and one with $m=-1$:
\begin{displaymath}
r\textrm{-mode } (A) \propto \textrm{e}^{-\textrm{i}\omega t} \quad \textrm{and} \quad \textrm{warp } (B)\propto \textrm{e}^{\textrm{i}\phi}
\end{displaymath}
\begin{displaymath}
 \Rightarrow\quad r\textrm{-mode }\times\textrm{ warp}\propto AB+AB^* \propto \textrm{e}^{\textrm{i}\phi-\textrm{i}\omega t}+\textrm{e}^{-\textrm{i}\phi-\textrm{i}\omega t}.
\end{displaymath}
Since we are only interested in the action of this coupling on the intermediate mode with $m=1$, because the one with $m=-1$ does not have any Lindblad resonances (location where the wave should be excited) in the disc, these forcing terms are projected on to $\textrm{e}^{\textrm{i}\phi-\textrm{i}\omega t}$. Similarly, the interaction of the intermediate mode with the warp gives rise to a $m=2$ mode in addition to the axisymmetric r~mode we are interested in:

\begin{displaymath}
\textrm{intermediate mode } (A) \propto \textrm{e}^{\textrm{i}\phi-\textrm{i}\omega t} \quad \textrm{and} \quad \textrm{warp } (B)\propto \textrm{e}^{\textrm{i}\phi}
\end{displaymath}
\begin{displaymath}
 \Rightarrow\quad \textrm{intermediate mode }\times\textrm{ warp}\propto AB+AB^* \propto \textrm{e}^{\textrm{i}2\phi-\textrm{i}\omega t}+\textrm{e}^{-\textrm{i}\omega t}.
\end{displaymath}
Therefore, these forcing terms should be projected on to $\textrm{e}^{-\textrm{i}\omega t}$, which means that complex conjugates of warp quantities will appear in the equations.

After separating variables, and projecting the forcing terms appropriately, the equations to be solved can be written as
\begin{eqnarray}
-\textrm{i}\omega u_{\textrm{R}r}=2\Omega u_{\textrm{R}\phi}-\frac{\textrm{d}h_\textrm{R}}{\textrm{d}r}-\frac{u_{\textrm{I}r}}{2}\frac{\textrm{d}u^*_{\textrm{W}r}}{\textrm{d}r}H+\textrm{i}u_{\textrm{I}\phi}\frac{u^*_{\textrm{W}r}}{2r}H+u_{\textrm{I}\phi}\frac{u^*_{\textrm{W}\phi}}{r}H-\frac{u^*_{\textrm{W}r}}{2}\frac{\textrm{d}u_{\textrm{I}r}}{\textrm{d}r}H-\textrm{i}u^*_{\textrm{W}\phi}\frac{u_{\textrm{I}r}}{2r}H
\label{fe1}
\end{eqnarray}
\begin{equation}
-\textrm{i}\omega u_{\textrm{R}\phi}=-\frac{\kappa^2}{2\Omega} u_{\textrm{R}r}-\frac{u_{\textrm{I}r}}{2}\frac{\textrm{d}u^*_{\textrm{W}\phi}}{\textrm{d}r}H-u_{\textrm{I}\phi}\frac{u^*_{\textrm{W}r}}{2r}H-\frac{u^*_{\textrm{W}r}}{2}\frac{\textrm{d}u_{\textrm{I}\phi}}{\textrm{d}r}H-u^*_{\textrm{W}\phi}\frac{u_{\textrm{I}r}}{2r}H
\end{equation}
\begin{equation}
-\textrm{i}\omega u_{\textrm{R}z}=-\frac{h_\textrm{R}}{H}-\frac{u_{\textrm{I}r}}{2}\frac{\textrm{d}u^*_{\textrm{W}z}}{\textrm{d}r}+\textrm{i}u_{\textrm{I}\phi}\frac{u^*_{\textrm{W}z}}{2r}
\end{equation}
\begin{equation}
-\textrm{i}\omega h_{\textrm{R}}=\Omega_z^2H u_{\textrm{R}z}-\frac{c_\mathrm{s}^2}{r}\frac{\textrm{d}(ru_{\textrm{R}r})}{\textrm{d}r}-\frac{u_{\textrm{I}r}}{2}\frac{\textrm{d}h^*_{\textrm{W}}}{\textrm{d}r}H+\textrm{i}u_{\textrm{I}\phi}\frac{h^*_{\textrm{W}}}{2r}H-\frac{u^*_{\textrm{W}r}}{2}\frac{\textrm{d}h_{\textrm{I}}}{\textrm{d}r}H-\textrm{i}u^*_{\textrm{W}\phi}\frac{h_{\textrm{I}}}{2r}H
\end{equation}
\begin{eqnarray}
-\textrm{i}\omega u_{\textrm{I}r}=-(\textrm{i}+\beta)\Omega u_{\textrm{I}r}+2\Omega u_{\textrm{I}\phi}-\frac{\textrm{d}h_\textrm{I}}{\textrm{d}r}-\frac{u_{\textrm{R}r}}{2}\frac{\textrm{d}u_{\textrm{W}r}}{\textrm{d}r}H-\textrm{i}u_{\textrm{R}\phi}\frac{u_{\textrm{W}r}}{2r}H+u_{\textrm{R}\phi}\frac{u_{\textrm{W}\phi}}{r}H-u_{\textrm{R}r}\frac{u_{\textrm{W}z}}{2H}-\frac{u_{\textrm{W}r}}{2}\frac{\textrm{d}u_{\textrm{R}r}}{\textrm{d}r}H\nonumber\\-u_{\textrm{R}z}\frac{u_{\textrm{W}r}}{2}
\end{eqnarray}
\begin{eqnarray}
-\textrm{i}\omega u_{\textrm{I}\phi}=-(\textrm{i}+\beta)\Omega u_{\textrm{I}\phi}-\frac{\kappa^2}{2\Omega} u_{\textrm{I}r}-\frac{\textrm{i}h_\textrm{I}}{r}-\frac{u_{\textrm{R}r}}{2}\frac{\textrm{d}u_{\textrm{W}\phi}}{\textrm{d}r}H-\textrm{i}u_{\textrm{R}\phi}\frac{u_{\textrm{W}\phi}}{2r}H-u_{\textrm{R}\phi}\frac{u_{\textrm{W}r}}{2r}H-\frac{u_{\textrm{W}r}}{2}\frac{\textrm{d}u_{\textrm{R}\phi}}{\textrm{d}r}H-u_{\textrm{W}\phi}\frac{u_{\textrm{R}r}}{2r}H\nonumber\\-u_{\textrm{R}z}\frac{u_{\textrm{W}\phi}}{2}-u_{\textrm{W}z}\frac{u_{\textrm{R}\phi}}{2H}
\end{eqnarray}
\begin{eqnarray}
-\textrm{i}\omega h_{\textrm{I}}=-(\textrm{i}+\beta)\Omega h_{\textrm{I}}-\frac{c_\mathrm{s}^2}{r}\frac{\textrm{d}(ru_{\textrm{I}r})}{\textrm{d}r}-c_\mathrm{s}^2\frac{\textrm{i}u_{\textrm{I}\phi}}{r}-\frac{u_{\textrm{R}r}}{2}\frac{\textrm{d}h_{\textrm{W}}}{\textrm{d}r}H-\textrm{i}u_{\textrm{R}\phi}\frac{h_{\textrm{W}}}{2r}H-\frac{u_{\textrm{W}r}}{2}\frac{\textrm{d}h_{\textrm{R}}}{\textrm{d}r}H-u_{\textrm{W}z}\frac{h_\textrm{R}}{2H}-u_{\textrm{R}z}\frac{h_\textrm{W}}{2}.
\label{fe2}
\end{eqnarray}
This system is linear in the unknowns for the r and intermediate modes. The warp, which couples these modes together, is assumed to be known.

For the interaction with the $n=2$ intermediate mode, similar equations need to be solved. After separating variables and projecting forcing terms appropriately, the equations describing this interaction read
\begin{eqnarray}
-\textrm{i}\omega u_{\textrm{R}r}=2\Omega u_{\textrm{R}\phi}-\frac{\textrm{d}h_\textrm{R}}{\textrm{d}r}-u_{\textrm{I}r}\frac{\textrm{d}u^*_{\textrm{W}r}}{\textrm{d}r}H+\textrm{i}u_{\textrm{I}\phi}\frac{u^*_{\textrm{W}r}}{r}H+2u_{\textrm{I}\phi}\frac{u^*_{\textrm{W}\phi}}{r}H-u^*_{\textrm{W}r}\frac{\textrm{d}u_{\textrm{I}r}}{\textrm{d}r}H-\textrm{i}u^*_{\textrm{W}\phi}\frac{u_{\textrm{I}r}}{r}H-\frac{u^*_{\textrm{W}z}}{H}u_{\textrm{I}r}+\frac{u^*_{\textrm{W}r}}{2}u_{\textrm{I}z}
\label{int1}
\end{eqnarray}
\begin{eqnarray}
-\textrm{i}\omega u_{\textrm{R}\phi}=-\frac{\kappa^2}{2\Omega} u_{\textrm{R}r}-u_{\textrm{I}r}\frac{\textrm{d}u^*_{\textrm{W}\phi}}{\textrm{d}r}H-u_{\textrm{I}\phi}\frac{u^*_{\textrm{W}r}}{r}H-u^*_{\textrm{W}r}\frac{\textrm{d}u_{\textrm{I}\phi}}{\textrm{d}r}H-u^*_{\textrm{W}\phi}\frac{u_{\textrm{I}r}}{r}H-\frac{u^*_{\textrm{W}z}}{H}u_{\textrm{I}\phi}+\frac{u^*_{\textrm{W}\phi}}{2}u_{\textrm{I}z}
\end{eqnarray}
\begin{equation}
-\textrm{i}\omega u_{\textrm{R}z}=-\frac{h_\textrm{R}}{H}-\frac{u^*_{\textrm{W}r}}{2}\frac{\textrm{d}u_{\textrm{I}z}}{\textrm{d}r}H-\textrm{i}u^*_{\textrm{W}\phi}\frac{u_{\textrm{I}z}}{2r}H-\frac{u^*_{\textrm{W}z}}{2H}u_{\textrm{I}z}
\end{equation}
\begin{eqnarray}
-\textrm{i}\omega h_{\textrm{R}}=\Omega_z^2H u_{\textrm{R}z}-\frac{c_\mathrm{s}^2}{r}\frac{\textrm{d}(ru_{\textrm{R}r})}{\textrm{d}r}-u_{\textrm{I}r}\frac{\textrm{d}h^*_{\textrm{W}}}{\textrm{d}r}H+\textrm{i}u_{\textrm{I}\phi}\frac{h^*_{\textrm{W}}}{r}H-u^*_{\textrm{W}r}\frac{\textrm{d}h_{\textrm{I}}}{\textrm{d}r}H-\textrm{i}u^*_{\textrm{W}\phi}\frac{h_{\textrm{I}}}{r}H-\frac{u^*_{\textrm{W}z}}{H}h_\textrm{I}-\frac{h^*_\textrm{W}}{2}u_{\textrm{I}z}
\end{eqnarray}
\begin{eqnarray}
-\textrm{i}\omega u_{\textrm{I}r}=-(\textrm{i}+\beta)\Omega u_{\textrm{I}r}+2\Omega u_{\textrm{I}\phi}-\frac{\textrm{d}h_\textrm{I}}{\textrm{d}r}-\frac{u_{\textrm{R}r}}{2}\frac{\textrm{d}u_{\textrm{W}r}}{\textrm{d}r}H-\textrm{i}u_{\textrm{R}\phi}\frac{u_{\textrm{W}r}}{2r}H+u_{\textrm{R}\phi}\frac{u_{\textrm{W}\phi}}{r}H-\frac{u_{\textrm{W}r}}{2}\frac{\textrm{d}u_{\textrm{R}r}}{\textrm{d}r}H
\end{eqnarray}
\begin{eqnarray}
-\textrm{i}\omega u_{\textrm{I}\phi}=-(\textrm{i}+\beta)\Omega u_{\textrm{I}\phi}-\frac{\kappa^2}{2\Omega} u_{\textrm{I}r}-\frac{\textrm{i}h_\textrm{I}}{r}-\frac{u_{\textrm{R}r}}{2}\frac{\textrm{d}u_{\textrm{W}\phi}}{\textrm{d}r}H-\textrm{i}u_{\textrm{R}\phi}\frac{u_{\textrm{W}\phi}}{2r}H-u_{\textrm{R}\phi}\frac{u_{\textrm{W}r}}{2r}H-\frac{u_{\textrm{W}r}}{2}\frac{\textrm{d}u_{\textrm{R}\phi}}{\textrm{d}r}H-u_{\textrm{W}\phi}\frac{u_{\textrm{R}r}}{2r}H
\end{eqnarray}
\begin{equation}
-\textrm{i}\omega u_{\textrm{I}z}=-(\textrm{i}+\beta)\Omega u_{\textrm{I}z}-\frac{2h_\textrm{I}}{H}-\frac{u_{\textrm{R}r}}{2}\frac{\textrm{d}u_{\textrm{W}z}}{\textrm{d}r}-i\frac{u_{\textrm{W}z}}{2r}u_{2\phi}-\frac{H}{2}u_{\textrm{W}r}\frac{\textrm{d}u_{\textrm{R}z}}{\textrm{d}r}
\end{equation}
\begin{eqnarray}
-\textrm{i}\omega h_{\textrm{I}}=-(\textrm{i}+\beta)\Omega h_{\textrm{I}}-\frac{c_\mathrm{s}^2}{r}\frac{\textrm{d}(ru_{\textrm{I}r})}{\textrm{d}r}-c_\mathrm{s}^2\frac{\textrm{i}u_{\textrm{I}\phi}}{r}+\Omega_z^2Hu_{\textrm{I}z}-\frac{u_{\textrm{R}r}}{2}\frac{\textrm{d}h_{\textrm{W}}}{\textrm{d}r}H-\textrm{i}u_{\textrm{R}\phi}\frac{h_{\textrm{W}}}{2r}H-\frac{u_{\textrm{W}r}}{2}\frac{\textrm{d}h_{\textrm{R}}}{\textrm{d}r}H.
\label{int2}
\end{eqnarray}
As before, the system is linear in the unknowns for the r and intermediate modes. It should be noted that although the same notation is used in the systems (\ref{fe1})--(\ref{fe2}) and (\ref{int1})--(\ref{int2}) to represent the intermediate mode quantities, they refer to two different modes: both with the same frequency and azimuthal mode number $m=1$, but with different vertical mode number ($n=0$ in the first system and $n=2$ in the second).

 \subsection{Wave coupling in an eccentric disc}
 
After separating variables, and projecting the forcing terms appropriately, as done above for the interactions in a warped disc, the equations describing the coupling between the trapped r~mode, eccentric disc and $n=1$ intermediate mode read
\begin{equation}
-\textrm{i}\omega u_{\textrm{R}r}=2\Omega u_{\textrm{R}\phi}-\frac{\textrm{d}h_\textrm{R}}{\textrm{d}r}-\frac{u_{\textrm{I}r}}{2}\frac{\textrm{d}u^*_{\textrm{E}r}}{\textrm{d}r}-\frac{u^*_{\textrm{E}r}}{2}\frac{\textrm{d}u_{\textrm{I}r}}{\textrm{d}r}-\frac{\textrm{i}u^*_{\textrm{E}\phi}}{2r}u_{\textrm{I}r}+\frac{u_{\textrm{I}\phi}}{2r}(\textrm{i}u^*_{\textrm{E}r}+u^*_{\textrm{E}\phi})
\label{inte1}
\end{equation}
\begin{equation}
-\textrm{i}\omega u_{\textrm{R}\phi}=-\frac{\kappa^2}{2\Omega} u_{\textrm{R}r}-\frac{u_{\textrm{I}r}}{2}\left(\frac{\textrm{d}u^*_{\textrm{E}\phi}}{\textrm{d}r}+\frac{u^*_{\textrm{E}\phi}}{r}\right)-\frac{u_{\textrm{I}\phi}}{2}\frac{u^*_{\textrm{E}r}}{r}-\frac{u^*_{\textrm{E}r}}{2}\frac{\textrm{d}u_{\textrm{I}r}}{\textrm{d}r}
\end{equation}
\begin{equation}
-\textrm{i}\omega u_{\textrm{R}z}=-\frac{h_\textrm{R}}{H}-\frac{\textrm{i}u_{\textrm{I}z}}{2}\frac{u^*_{\textrm{E}\phi}}{r}-\frac{u^*_{\textrm{E}r}}{2}\frac{\textrm{d}u_{\textrm{I}z}}{\textrm{d}r}
\end{equation}
\begin{equation}
-\textrm{i}\omega h_{\textrm{R}}=\Omega_z^2H u_{\textrm{R}z}-\frac{c_\mathrm{s}^2}{r}\frac{\textrm{d}(ru_{\textrm{R}r})}{\textrm{d}r}-\frac{u_{\textrm{I}r}}{2}\frac{\textrm{d}h^*_\textrm{E}}{\textrm{d}r}+\frac{\textrm{i}u_{\textrm{I}\phi}}{2r}\frac{h^*_\textrm{E}}{r}-\frac{u^*_{\textrm{E}r}}{2}\frac{\textrm{d}h_\textrm{I}}{\textrm{d}r}-\frac{\textrm{i}h_\textrm{I}}{2}\frac{u^*_{\textrm{E}\phi}}{r}
\end{equation}
\begin{equation}
-\textrm{i}\omega u_{\textrm{I}r}=-(\textrm{i}+\beta)\Omega u_{\textrm{I}r}+2\Omega u_{\textrm{I}\phi}-\frac{\textrm{d}h_\textrm{I}}{\textrm{d}r}-\frac{u_{\textrm{R}r}}{2}\frac{\textrm{d}u_{\textrm{E}r}}{\textrm{d}r}-\frac{u_{\textrm{E}r}}{2}\frac{\textrm{d}u_{\textrm{R}r}}{\textrm{d}r}-\frac{u_{\textrm{R}\phi}}{2r}(\textrm{i}u_{\textrm{E}r}-u_{\textrm{E}\phi})
\end{equation}
\begin{equation}
-\textrm{i}\omega u_{\textrm{I}\phi}=-(\textrm{i}+\beta)\Omega u_{\textrm{I}\phi}-\frac{\kappa^2}{2\Omega} u_{\textrm{I}r}-\frac{\textrm{i}h_\textrm{I}}{r}-\frac{u_{\textrm{R}r}}{2}\frac{\textrm{d}u_{\textrm{E}\phi}}{\textrm{d}r}-\frac{u_{\textrm{R}r}}{2}\frac{u_{\textrm{E}\phi}}{r}-\frac{u_{\textrm{R}\phi}}{2r}(\textrm{i}u_{\textrm{E}\phi}+u_{\textrm{E}r})-\frac{u_{\textrm{E}r}}{2}\frac{\textrm{d}u_{\textrm{R}\phi}}{\textrm{d}r}
\end{equation}
\begin{equation}
-\textrm{i}\omega u_{\textrm{I}z}=-(\textrm{i}+\beta)\Omega u_{\textrm{I}z}-\frac{h_\textrm{I}}{H}-\frac{u_{\textrm{E}r}}{2}\frac{\textrm{d}u_{\textrm{R}z}}{\textrm{d}r}
\end{equation}
\begin{equation}
-\textrm{i}\omega h_{\textrm{I}}=-(\textrm{i}+\beta)\Omega h_{\textrm{I}}-\frac{c_\mathrm{s}^2}{r}\frac{\textrm{d}(ru_{\textrm{I}r})}{\textrm{d}r}-c_\mathrm{s}^2\frac{\textrm{i}u_{\textrm{I}\phi}}{r}+\Omega_z^2Hu_{\textrm{I}z}-\frac{u_{\textrm{R}r}}{2}\frac{\textrm{d}h_{\textrm{E}}}{\textrm{d}r}-\textrm{i}u_{\textrm{R}\phi}\frac{h_{\textrm{E}}}{2r}-\frac{u_{\textrm{E}r}}{2}\frac{\textrm{d}h_{\textrm{R}}}{\textrm{d}r}.
\label{inte2}
\end{equation}

\bsp

\label{lastpage}

\end{document}